\documentclass[aps,twocolumn,a4paper,pra]{revtex4}
\usepackage{psfig}
\usepackage{amssymb}

\newcommand{\net}[1]{\{\mathrm{#1}\}}
\newcommand{\netsimple}[1]{_\mathrm{#1}}

\begin{document}

\title{The Casimir force and the quantum theory of lossy optical cavities}
\author{Cyriaque Genet}
\altaffiliation[Present address: ]{Huygens Laboratory, Leiden University, 
P.O. Box 9504, 2300 RA Leiden The Netherlands} 
\author{Astrid Lambrecht}
\author{Serge Reynaud}
\email[]{reynaud@spectro.jussieu.fr}
\homepage[]{www.spectro.jussieu.fr/Vacuum}
\affiliation{Laboratoire Kastler Brossel 
\footnote{Unit\'{e} de l'Ecole Normale Sup\'{e}rieure, de l'Universit\'{e} Pierre
et Marie Curie, et du Centre National de la Recherche Scientifique}, 
case 74, Campus Jussieu, F-75252 Paris France}
\date{February 10, 2003}

\begin{abstract}
We present a new derivation of the Casimir force between 
two parallel plane mirrors at zero temperature.
The two mirrors and the cavity they enclose
are treated as quantum optical networks. 
They are in general lossy and characterized 
by frequency dependent reflection amplitudes. 
The additional fluctuations accompanying losses are deduced 
from expressions of the optical theorem. 
A general proof is given for the theorem relating the spectral density 
inside the cavity to the reflection amplitudes seen by the inner fields. 
This density determines the vacuum radiation pressure and, 
therefore, the Casimir force. 
The force is obtained as an integral over the real frequencies, including 
the contribution of evanescent waves besides that of ordinary waves, and, 
then, as an integral over imaginary frequencies. 
The demonstration relies only on general properties obeyed by real mirrors
which also enforce general constraints for the variation of the 
Casimir force.
\end{abstract}
\maketitle

\section{Introduction}

An important prediction of quantum theory is the existence of irreducible
fluctuations of electromagnetic fields in vacuum. Besides
their numerous observable consequences in microscopic physics, vacuum 
fluctuations also have observable effects in macroscopic physics,
for example the Casimir force they exert on mirrors \cite{Casimir48}.  

Casimir calculated this force in a geometrical configuration where two plane 
mirrors are placed a distance $L$ apart and parallel to each other, 
the area $A$ of the mirrors being much larger than the squared distance 
$A \gg L^2$. He considered the ideal case of perfectly reflecting mirrors and 
obtained an expression which, remarkably, depends only on the 
geometrical quantities $A$ and $L$ and on the fundamental constants $\hbar$ and $c$ 
\begin{equation}
F_\mathrm{Cas}  =  \frac{\hbar c \pi^2 A}{240 L^4}  
\label{eqCasimir}
\end{equation}

This attractive force has been observed in a number of `historical' experiments 
\cite{Deriagin57,Spaarnay58,Tabor68,Black68,Sabisky73} 
which confirmed its existence and main properties 
\cite{Sparnaay89,Milonni94,Mostepanenko97}.
Several recent experiments reached an accuracy in the \% range by measuring
the force between a plane and a sphere \cite{Lamoreaux97,Mohideen98,Roy99,Harris00}
or two cylinders \cite{Ederth00}.
Similar experiments were also performed with MEMS \cite{Chan01s,Chan01p}
(see also \cite{Buks01}).
An experiment studied the plane-plane configuration considered by Casimir
\cite{Bressi02} but, as a consequence of the difficulties
associated with this geometry, reached only a 15\% accuracy
(see reviews of recent experiments in \cite{Bordag01,Lambrecht02}).

The Casimir force is the most accessible experimental consequence of 
vacuum fluctuations in the macroscopic world while vacuum energy is known to 
raise a serious problem with respect to gravity and cosmology
(see references in \cite{Reynaud01,Genet02}). This is a reason for testing 
the predictions of Quantum Field Theory concerning the Casimir effect
with the greatest care and accuracy. 
The theory of the Casimir force is also a key point for the experiments searching 
for the new weak forces predicted by theoretical unification models 
to arise at distances between nanometer and millimeter
\cite{Carugno97,Fischbach98,Bordag99,Fischbach99,Long99,Hoyle01,Adelberger02,Long02}.
The Casimir force is indeed the dominant effect between two neutral objects at $\mu$m
or sub$\mu$m distances so that an accurate knowledge of its theoretical expectation 
is as crucial as the precision of measurements in such experiments \cite{Lambrecht00}. 

In this context, it is essential to account for the differences between the 
ideal case considered by Casimir and the real experimental situation. 
Recent experiments use metallic mirrors which show perfect 
reflection only at frequencies below their plasma frequency. 
They are performed at room temperature, with the effect of 
thermal fluctuations superimposed to that of vacuum fluctuations. 
In the most accurate experiments, the force is measured between a plane and a 
sphere, and not between two parallel planes. The surface state of 
the plates, in particular their roughness, should also affect the force. 
A large number of works have been devoted to the study of these effects and we refer
the reader to \cite{Bordag01,Lambrecht02} for a bibliography.

The evaluation of the Casimir force between imperfect lossy mirrors at non zero 
temperature has given rise to a burst of controversial results 
\cite{Bostrom00,Svetovoy00,Bordag00,Lamoreaux01c,Sernelius01r,Sernelius01c,%
Bordag01r,Klimchitskaya01,Bezerra02,Lamoreaux02}
which constitutes a part of the motivations for the present work.
For the sake of comparing experimental measurements and theoretical expectations, 
it is necessary to have at one's disposal a reliable expression 
of the Casimir force in the experimental situation.
In the present paper, we focus our attention on the effect of imperfect 
reflection of the mirrors. Other effects, in particular the effect of temperature,
will be addressed in follow-on papers. 

We consider the original Casimir geometry with two perfectly plane and parallel 
mirrors. Except for these assumptions, we consider arbitrary frequency dependences
for the mirrors which, in particular, may be lossy.
We evaluate the Casimir force as the effect of vacuum radiation
pressure on the Fabry-Perot cavity formed by the two mirrors. 
The net force results from the balance between the repulsive and 
attractive contributions associated respectively with 
resonant or antiresonant frequencies. 
It is obtained as an integral over the axis of real frequencies, 
including the contribution of evanescent waves besides that of ordinary waves. 
It is then transformed into an integral over imaginary frequencies by using 
physical properties fulfilled by all real mirrors.

The formula obtained here for the Casimir force turns out to be identical to the 
expression already published in \cite{Jaekel91} but the new derivation has a wider 
scope of validity than the previous one since it remains valid for lossy mirrors. 
The fact that the formula keeps the same form despite the widening of the 
assumptions is intimately related to a theorem which relates the spectral density 
of the fields inside the cavity to the reflection amplitudes seen by the same fields. 
This theorem was demonstrated in \cite{Jaekel91} and \cite{Barnett98} in specific cases 
and we prove it in the present paper without any restriction. 
To this aim, we introduce a systematic treatment of lossy mirrors and cavities 
as dissipative networks \cite{Meixner}.
We define scattering and transfer matrices for elementary networks like the interface 
between two media or the propagation over a given length in a medium.
We then deduce the matrices associated with composed networks, like the optical 
slab or the multilayer mirror. 

The results obtained in this manner are therefore applicable to a large variety 
of mirrors, still with the assumption of perfect plane geometry. 
In the particular case of a slab with a large width, the Lifshitz expression 
\cite{Lifshitz56,LLcasimir} is recovered. At the limit of perfectly reflectors, 
the ideal Casimir formula (\ref{eqCasimir}) is obtained.
More generally, the expression gives the Casimir force as an integral written
in terms of the reflection amplitudes characterizing the two mirrors. 
This integral is finite as soon as the amplitudes obey the general properties
of scattering theory already alluded to.
In other words, the difficulties usually associated with the infiniteness of 
vacuum energy are solved by using the properties of real mirrors themselves 
rather than through an additional formal regularization technique. 

We finally show that the same physical properties constrain the variation of the Casimir 
force. In particular, they invalidate proposals which has been done for `tayloring' the force 
at will by using mirrors with specially designed scattering amplitudes \cite{Iacopini93,Ford93}. 
In these proposals, the balance between attractive and repulsive contributions to the force
is change, leading to the hope that the Casimir force could reach large or 
have its sign changed from an attractive force to a repulsive one \cite{Iacopini93}.
Using the simple model of a one-dimensional space, it has already been shown \cite{Lambrecht97} 
that these hopes cannot be met for arbitrary mirrors built up with dielectric layers. 
Here, the argument is generalized to the Casimir geometry in three-dimensional space
with the following conclusions~: the Casimir force cannot exceed the value 
obtained for perfect mirrors, it remains attractive for any cavity length 
and its value is a decreasing function of the cavity length. 
This is true for any mirror obtained by piling up layers of media described by dielectric 
functions. This definition of multilayer dielectric mirrors includes the case of metallic 
layers, provided that magnetic effects play a negligible role in the optical response.

\section{Vacuum field modes}

As explained in the Introduction, we consider in this paper the original 
Casimir geometry with perfectly plane and parallel mirrors aligned
along the directions $x$ and $y$. This configuration obeys a 
symmetry with respect to time translation as well as transverse space 
translations along these directions. We use bold letters for 
two-dimensional vectors along these directions and denote 
$\mathbf{r} \equiv \left( x,y\right)$ the transverse position. 
As a consequence of this symmetry, the frequency $\omega $, the
transverse vector $\mathbf{k} \equiv \left( k_{x},k_{y}\right)$ and the 
polarization $p=\mathrm{TE},\mathrm{TM}$ are preserved throughout the
scattering processes on a mirror or a cavity. The scattering couples 
only the free vacuum modes which have the same values for the preserved quantum 
numbers and differ by the sign of the longitudinal component $k_{z}$ of the wavevector.

In the present section, we introduce notations for the vacuum field modes,
first in empty space and then in a dielectric medium. These notations are
chosen to be well-adapted to the symmetry of the problem.

\subsection{Vacuum modes in empty space}

In empty space, the components of the wavevector are given for each field mode
by the frequency $\omega$, the incidence angle $\theta$ and the azimuthal 
angle $\varphi $  
\begin{eqnarray}
k_{x}= |\mathbf{k}| \cos \varphi &&\qquad 
|\mathbf{k}| = \frac\omega{c} \sin\theta \nonumber \\
k_{y}= |\mathbf{k}| \sin \varphi &&\qquad 
k_{z}= \frac{\omega }{c} \cos \theta
\label{dispersionrelation}
\end{eqnarray}
$|\mathbf{k}|$ is the modulus of the transverse wavevector and 
the longitudinal component $k_{z}$ may be expressed in terms of 
the preserved quantities $\omega$ and $\mathbf{k}$ 
\begin{equation}
k_{z}=\phi \sqrt{\frac{\omega ^{2}}{c^{2}}-\mathbf{k}^{2}}\qquad \phi =\pm 1
\end{equation}
$\phi$ is defined as the sign of $\cos\theta$ and represents the direction of 
propagation with $+1$ and $-1$ corresponding respectively to rightward 
and leftward propagation.

The two polarizations $p=\mathrm{TE},\mathrm{TM}$ are defined by the transversality 
with the incidence plane of electric and magnetic fields respectively.
They are given by the unit electric vectors $\widehat{\epsilon}$ 
\begin{eqnarray}
\widehat{\epsilon}_{x}^\mathrm{TM} = \cos \theta \cos \varphi &&\qquad
\widehat{\epsilon}_{x}^\mathrm{TE} = -\sin \varphi \nonumber \\
\widehat{\epsilon}_{y}^\mathrm{TM} = \cos \theta \sin \varphi &&\qquad 
\widehat{\epsilon}_{y}^\mathrm{TE} = \cos \varphi \nonumber \\
\widehat{\epsilon}_{z}^\mathrm{TM} = -\sin \theta &&\qquad 
\widehat{\epsilon}_{z}^\mathrm{TE} = 0  
\label{Epolar}
\end{eqnarray}
or, equivalently, the unit magnetic vectors 
$\widehat{\beta}^\mathrm{TM} =\widehat{\epsilon}^\mathrm{TE}$ and 
$\widehat{\beta}^\mathrm{TE}=-\widehat{\epsilon}^\mathrm{TM}$. 
For each mode, the wavevector and polarization
vectors form an orthogonal spatial basis. 
We have chosen linear polarizations described by real components; hence the unit 
vectors $\widehat{\epsilon }$ and $\widehat{\beta}$ are not affected by the complex 
conjugation appearing below in the relation between positive and negative frequencies.

The two modes corresponding to the same values of $\omega$, $\mathbf{k}$ and 
$p$ but opposite values of $\phi$ are coupled by scattering on a mirror. 
For this reason, we introduce a label $m\equiv \left( \omega ,\mathbf{k},p\right)$ 
gathering the values of $\omega $, $\mathbf{k}$ and $p$.
A mode freely propagating in vacuum is thus labeled by $m$ and $\phi $ and the
summation over modes is described by the symbols 
\begin{eqnarray}
\sum_{m\phi } &\equiv& \sum_{p}\int \frac{\mathrm{d}^{2}\mathbf{k}}{4\pi ^{2}}
\int_{-\infty }^{\infty }\frac{\mathrm{d}k_{z}}{2\pi } \nonumber \\
&\equiv& \sum_{\phi}\sum_{p}\int \frac{\mathrm{d}^{2}\mathbf{k}}{4\pi ^{2}}
\int_{0}^{\infty }\frac{\omega }{ck_{z}}\frac{\mathrm{d}\omega }{2\pi c} 
\label{mphi}
\end{eqnarray}
Note that $\phi $ appears implicitly as the sign of $k_{z}$ in the
first form whereas it appears explicitly in the second one.

\begin{widetext}
The free vacuum fields are then written as linear superpositions of modes 
\begin{eqnarray}
E\left( \mathbf{r},z,t\right) &=&\sqrt{cZ_\mathrm{vac}}\sum_{m\phi }\ 
\sqrt{\frac{\hbar \omega }{2}}\widehat{\epsilon }_{m}^{\phi }\left( e_{m}^{\phi }\
e^{-i\left( \omega t-\mathbf{k.r}-k_{z}z\right) }+\left( e_{m}^{\phi }\right)
^\dagger \ e^{i\left( \omega t-\mathbf{k.r}-k_{z}z\right) }\right)  \nonumber \\
B\left( \mathbf{r},z,t\right) &=&\sqrt{\frac{Z_\mathrm{vac}}{c}}\sum_{m\phi }\ 
\sqrt{\frac{\hbar \omega }{2}}\widehat{\beta }_{m}^{\phi }\left( e_{m}^{\phi
}\ e^{-i\left( \omega t-\mathbf{k.r}-k_{z}z\right) }+\left( e_{m}^{\phi
}\right) ^\dagger \ e^{i\left( \omega t-\mathbf{k.r}-k_{z}z\right) }\right)
\label{EBfields}
\end{eqnarray}
The vacuum impedance $Z_\mathrm{vac}= \mu _{0}c \simeq 377\Omega$
describes the electromagnetic constants in vacuum.
In the following, the symbol $\varepsilon $ will be reserved to the relative
permittivity with the value 1 in vacuum.
\end{widetext}

The quantum field amplitudes $e_{m}^{\phi}$ and $\left( e_{m}^{\phi}\right)^\dagger $ 
correspond to positive and negative frequency components. They fit 
the definition of annihilation and creation operators of quantum field theory and
obey the canonical commutation relations \cite{CCT87} 
\begin{eqnarray}
\left[ e_{m^\prime }^{\phi ^\prime },e_{m}^{\phi}{}^\dagger \right] &=&
\left( 2\pi \right) ^{3}\delta ^{(2)}\left( \mathbf{k-k}^{\prime
}\right) \delta \left( k_{z}-k_{z}^\prime \right) \delta _{pp^{\prime
}}\delta _{\phi \phi ^\prime }  \nonumber \\
&\equiv& \delta _{mm^\prime }\delta _{\phi \phi ^{\prime}} \nonumber \\
\left[ e_{m^\prime }^{\phi ^\prime },e_{m}^{\phi}\right] &=&
\left[ e_{m^\prime }^{\phi ^\prime }{}^{\dagger},e_{m}^{\phi}{}^\dagger \right] =0  
\label{commut}
\end{eqnarray}
In the vacuum state, the anticommutators of quantum amplitudes are derived 
from the corresponding commutators 
\begin{eqnarray}
\left\langle e_{m^\prime }^{\phi ^\prime } \cdot e_{m}^{\phi}{}^\dagger \right\rangle 
_{\mathrm{vac}} &=& 
\frac{1}{2} \left[ e_{m^\prime }^{\phi ^\prime }, e_{m}^{\phi}{}^\dagger \right] 
= \frac{1}{2} \delta _{mm^\prime }\delta _{\phi \phi ^{\prime}} \nonumber \\
\left\langle e_{m^\prime }^{\phi ^\prime } \cdot e_{m}^{\phi} \right\rangle 
_{\mathrm{vac}} &=& 
\frac{1}{2} \left[ e_{m^\prime }^{\phi ^\prime }, e_{m}^{\phi}{} \right] = 0
\label{anticommVacuum}
\end{eqnarray}
The dot symbol represents a symmetrized product.

\subsection{Stress tensor in empty space}

The energy density per unit volume $T_{00}$ is a quadratic form
of the fields $E$ and $B$
\begin{equation}
T_{00} \left( \mathbf{r},z,t\right) = \frac{1}{2cZ_\mathrm{vac}}
\left( E^{2} + c^{2} B^{2} \right) 
\end{equation} 
When subsituting the expression of free fields, $T_{00}$ is obtained as a 
bilinear form of the field amplitudes. 
Here, we study the averaged radiation pressure in the vacuum state which
leads to a contraction $m^\prime=m$ in the sums over modes. 
Using the vacuum property (\ref{anticommVacuum}), we find the averaged energy density in 
vacuum equal to the sum over the modes of $\frac{\hbar \omega }{2}$ 
\begin{equation}
\left\langle T_{00} \left( \mathbf{r},z,t\right) \right\rangle _\mathrm{vac}=
\sum_{m\phi }\ \frac{\hbar\omega }{2}
\label{T00Vac}
\end{equation} 
As it is well-known, this energy density is infinite.

The radiation pressure on plane mirrors oriented along $xy$ directions is 
determined by the component $T_{zz}$ of the Maxwell stress tensor 
\begin{equation}
T_{zz}\left( \mathbf{r},z, t\right) = \frac{1}{2Z_\mathrm{vac}}
\left( E \cdot \overline{E} + c^{2} B \cdot \overline{B} \right)
\end{equation} 
Here, the dot symbol represents a symmetrized product of the quantum amplitudes 
and, simultaneously, a scalar product of the vectors; the overline symbol 
describes the mathematical reflexion of a vector with respect to the plane $xy$ 
\begin{equation}
\overline{E}_{x}= E_x \qquad \overline{E}_{y}= E_y \qquad 
\overline{E}_{z}= - E_z 
\end{equation}

As for $T_{00}$, averaging $T_{zz}$ in vacuum state leads to a contraction
over the modes with the result 
\begin{eqnarray}
\left\langle T_{zz} \left( \mathbf{r},z,t\right) \right\rangle _\mathrm{vac} 
&=& \sum_{m\phi}\frac{\hbar\omega }{4} \left( 
\widehat{\epsilon}_m^\phi . \overline{\widehat{\epsilon}_m^\phi}
+ \widehat{\beta}_m^\phi . \overline{\widehat{\beta}_m^\phi}
\right) \nonumber \\
&=&\sum_{m\phi }\ \frac{\hbar \omega }{2} \cos ^{2}\theta  
\label{TzzVac}
\end{eqnarray}
This expression is similar to the expression (\ref{T00Vac}) of the energy
density with an extra factor $\cos ^{2}\theta$ well-known in studies of 
radiation pressure.
The sum over modes is still infinite but this infiniteness problem will be solved
in the forthcoming calculation of the Casimir force.

\subsection{Fields in dielectric media}

In the following, we consider mirrors built up as dielectric multilayers.
Each dielectric medium is characterized by a relative permittivity 
$\varepsilon \left[ \omega \right]$ or, equivalently, an index of refraction 
$n\left[ \omega \right] = \sqrt{\varepsilon \left[ \omega \right]}$
depending on frequency. The magnetic permeability is 
kept equal to its vacuum value since this corresponds to all experimental situations
studied so far. We stress again that this definition of dielectric mirrors includes 
the case of metals as long as the magnetic response plays a negligible role. 
We consider layers thick enough so that the dielectric response is local, i.e. described 
by a wavevector-independent permittivity $\varepsilon \left[ \omega \right]$. 

We will sometimes take the plasma model as a first
description of metallic optical response 
\begin{equation}
\varepsilon\left[ \omega \right] = 1 - \frac{\omega_\mathrm{P}^2}{\omega^2}
\qquad  \omega_\mathrm{P} = \frac{2\pi c}{\lambda_\mathrm{P}}
\label{PlasmaModel}
\end{equation}
where $\omega_\mathrm{P}$ and $\lambda_\mathrm{P}$ represent respectively the
plasma frequency and the plasma wavelength. 
This simple model is not sufficient for an accurate evaluation of the Casimir force
between real mirrors \cite{Lambrecht00}. 
To this aim, it is necessary to describe the optical response of metals
with a dissipative part associated with electronic relaxation processes.
As a consequence of causality, the real and imaginary parts of $n$ 
are related to each other through the Kramers-Kronig dispersion relations 
\cite{LLcausality}.

For any function of frequency more generally, 
causality is unambiguously characterized in terms of analyticity properties~: 
$n\left[ \omega \right]$ or $\varepsilon \left[ \omega \right] $ 
are analytical functions of $\omega $ in the `physical domain' of the complex frequency 
plane, that is the domain of frequencies $\omega$ with a positive imaginary part 
$\Im \omega >0$. 
This property is obeyed by other response functions to be encountered below 
and it will play an important role in the derivation of the Casimir force. 
We will introduce an equivalent notation $\xi$ for complex frequencies 
with the physical domain now defined by a positive real part for $\xi$ 
\begin{equation}
\omega \equiv i\xi \qquad \Re \xi >0
\end{equation}

The dispersion relation (\ref{dispersionrelation}) is changed
inside a refractive medium to 
\begin{eqnarray}
k_{x}= |\mathbf{k}| \cos \varphi &&\qquad 
|\mathbf{k}| = n \left[ \omega \right] \frac\omega{c} \sin\theta \nonumber \\
k_{y}= |\mathbf{k}| \sin \varphi &&\qquad 
k_{z}= n \left[ \omega \right] \frac{\omega }{c} \cos \theta
\label{dispersionrelationIndex}
\end{eqnarray}
The preservation of $\omega$ and $\mathbf{k}$ at the traversal of an interface
is equivalent to the Snell-Descartes law of refraction. 

The sign has to be carefully chosen when extracting the square root 
to express $k_{z}$ in terms of the conserved quantities $\omega $
and $\mathbf{k}$. As soon as the refractive index contains an imaginary part,
this is also the case for $k_{z}$ and the dephasing $\exp\left( ik_{z} z\right)$ 
associated with propagation includes an extinction factor. In order to ensure that 
this factor is effectively a decreasing exponential, we have to choose a specific root 
defined differently for the two propagation directions $\phi=\pm 1$ of the field 
\begin{eqnarray}
&&k_{z}\equiv i \phi \kappa \nonumber \\
\kappa &=&\sqrt{\varepsilon \left[ i\xi \right] \frac{\xi ^{2}}{c^{2}}+\mathbf{k}^{2}}
\qquad \Re\kappa >0
\label{dispersionKappa}
\end{eqnarray}
The argument has been presented for freely propagating modes but it holds as
well for evanescent waves confined to the vicinity of an interface between
two media. In this case, the sign of $k_{z}$ is also chosen so that it corresponds 
to an extinction when the distance to the interface increases and this choice is 
still described by equation (\ref{dispersionKappa}). 
In the following, we will use systematically the notations $\xi$ and 
$\kappa$, keeping in mind that the causality relations have to be written 
for each value of the conserved quantity $\mathbf{k}$.  

Besides the dispersion relation (\ref{dispersionrelationIndex}), the dielectric
medium also changes the impedance, that is the ratio between magnetic and electric 
field amplitudes. Precisely, the impedance is changed from the value $Z_\mathrm{vac}$ 
in empty space to the value $\frac{Z_\mathrm{vac}}n$ in a dielectric medium of index $n$, 
resulting in reflection at the interface.

\section{Mirrors as optical networks}

We now introduce the description of mirrors as optical networks. 
We present the scattering and transfer representations and the
relations between them. The transfer approach is well adapted to the
composition of networks which are piled up. We first consider elementary 
networks such as an interface or propagation inside a refractive medium. 
We then use the composition law to study composed networks 
such as the slab and multilayer.
In the present section, we only consider classical fields 
or, equivalently, mean quantum fields. 
The next section will be devoted to the full quantum treatment including 
the addition of noise associated with the losses inside the mirror.

\subsection{Scattering and transfer representations}

\begin{figure}[tbh]
\centerline{\psfig{figure=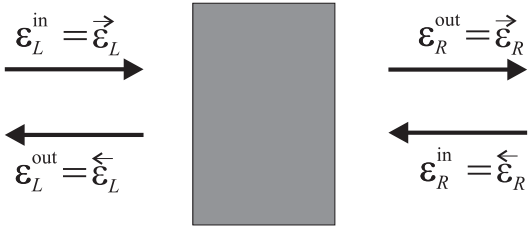,height=3cm}}
\caption{\label{FigNetwork}Scattering and transfer representations of a network.}
\end{figure}

We first introduce the scattering and transfer representations for an 
arbitrary network represented with two ports and four fields.
These fields are identified as lefthand/righthand (symbols `L' and `R'), 
rightward/leftward (arrows $\rightarrow$ and $\leftarrow$) or
input/output fields (labels `in' and `out'), as shown on Figure \ref{FigNetwork}.

\vspace*{4mm}\vspace*{4mm}

Let us emphasize that the arrows are a symbolic representation of the two modes 
coupled by the network which correspond to the same label $m$ and to the two
opposite signs $\phi = \pm 1$. The geometrical directions of propagation are 
given by the wavevectors of equation (\ref{dispersionrelationIndex}). The coupling
between the fields is described by reflection and transmission amplitudes 
represented below by scattering or transfer matrices. 

In the scattering point of view, we gather the input and output fields in twofold 
columns related by a $S-$matrix 
\begin{eqnarray}
&&\left| \mathcal{E}^{\mathrm{in}}\right\rangle =\left( 
\begin{tabular}{l}
$\mathcal{E}_{\mathrm{L}}^{\mathrm{in}}$ \\ 
$\mathcal{E}_{\mathrm{R}}^{\mathrm{in}}$%
\end{tabular}
\right) \qquad 
 \left| \mathcal{E}^{\mathrm{out}}\right\rangle =\left( 
\begin{tabular}{l}
$\mathcal{E}_{\mathrm{L}}^{\mathrm{out}}$ \\ 
$\mathcal{E}_{\mathrm{R}}^{\mathrm{out}}$%
\end{tabular}
\right) 
\nonumber \\
&&\left| \mathcal{E}^{\mathrm{out}}\right\rangle =S\left| 
\mathcal{E}^{\mathrm{in}}\right\rangle \qquad 
S=\left( 
\begin{tabular}{ll}
$r$ & $\overline{t}$ \\ 
$t$ & $\overline{r}$%
\end{tabular}
\right)  
\label{Smatrix}
\end{eqnarray}
$r$ and $\overline{r}$ are the reflection amplitudes while 
$t$ and $\overline{t}$ are the transmission amplitudes. 
We will also use an equivalent convention where the output ket 
is defined with the upper and lower components exchanged
\begin{eqnarray}
&&\widetilde{\left| \mathcal{E}^{\mathrm{out}}\right\rangle }
=\left( 
\begin{tabular}{l}
$\mathcal{E}^{\mathrm{out}}_\mathrm{R}$ \\ 
$\mathcal{E}^{\mathrm{out}}_\mathrm{L}$%
\end{tabular}
\right) = \eta \left| \mathcal{E}^{\mathrm{out}}\right\rangle  \qquad
\eta =\left( 
\begin{tabular}{ll}
$0$ & $1$ \\ 
$1$ & $0$%
\end{tabular}
\right)
\nonumber \\
&&\widetilde{\left| \mathcal{E}^{\mathrm{out}}\right\rangle }= 
\widetilde{S}\left| \mathcal{E}^{\mathrm{in}}\right\rangle \qquad 
\widetilde{S}=\eta S=\left( 
\begin{tabular}{ll}
$t$ & $\overline{r}$ \\ 
$r$ & $\overline{t}$%
\end{tabular}
\right)  
\label{SmatrixTilde}
\end{eqnarray}
This convention simplifies some algebraic manipulations 
while being completely equivalent to the former convention.
For comparison with previous works, note that the former notation 
(\ref{Smatrix}) was used in \cite{Jaekel91} whereas the latter one 
(\ref{SmatrixTilde}) was used in \cite{Lambrecht97}.

In the transfer point of view, the network is described by lefthand and righthand 
columns related by a $T-$matrix 
\begin{eqnarray}
&&\left| \mathcal{E}_{\mathrm{L}}\right\rangle =\left( 
\begin{tabular}{l}
$\mathcal{E}_{\mathrm{L}}^{\rightarrow }$ \\ 
$\mathcal{E}_{\mathrm{L}}^{\leftarrow }$%
\end{tabular}
\right) 
\qquad 
\left| \mathcal{E}_{\mathrm{R}}\right\rangle =\left( 
\begin{tabular}{l}
$\mathcal{E}_{\mathrm{R}}^{\rightarrow }$ \\ 
$\mathcal{E}_{\mathrm{R}}^{\leftarrow }$%
\end{tabular}
\right) 
\nonumber \\
&&\left| \mathcal{E}_{\mathrm{L}}\right\rangle =T\left| 
\mathcal{E}_{\mathrm{R}}\right\rangle \qquad T=\left( 
\begin{tabular}{ll}
$a$ & $b$ \\ 
$c$ & $d$%
\end{tabular}
\right)  \label{Tmatrix}
\end{eqnarray}

The matrix $\eta$ introduced in (\ref{SmatrixTilde}) exchanges 
the two directions of propagation. We also use in the following the matrices 
$\pi_\pm$ which project onto each direction  
\begin{equation}
\pi _{+} =\left( 
\begin{tabular}{ll}
$1$ & $0$ \\ 
$0$ & $0$%
\end{tabular}
\right) \qquad 
\pi _{-} =\left( 
\begin{tabular}{ll}
$0$ & $0$ \\ 
$0$ & $1$%
\end{tabular}
\right)
\end{equation}
These matrices obey simple rules which define an algebraic calculus in the space 
$\mathcal{M}_{2}\left( \mathbb{C}\right) $ of $2 \times 2$ matrices 
with complex coefficients
\begin{eqnarray}
\pi _{+}^{2} = \pi _{+} &&\qquad \pi _{-}^{2}=\pi _{-}\qquad 
\pi _{+}\pi _{-}=\pi _{-}\pi _{+}=0 \nonumber \\
\eta ^{2} = I &&\qquad \eta \pi _{+}=\pi _{-}\eta \qquad 
\eta \pi _{-}=\pi _{+}\eta 
\end{eqnarray}

The identification of Figure (\ref{FigNetwork}) is written as 
\begin{eqnarray}
\pi _{+}\left| \mathcal{E}_{\mathrm{R}}\right\rangle &=&\pi _{+}\widetilde{\left| 
\mathcal{E}^{\mathrm{out}}\right\rangle }\qquad \pi _{-}\left| 
\mathcal{E}_{\mathrm{R}}\right\rangle
=\pi _{-}\left| \mathcal{E}^{\mathrm{in}}\right\rangle  \nonumber \\
\pi _{+}\left| \mathcal{E}_{\mathrm{L}}\right\rangle &=&\pi _{+}\left| 
\mathcal{E}^{\mathrm{in}} \right\rangle \qquad \pi _{-}\left| 
\mathcal{E}_{\mathrm{L}}\right\rangle =\pi_{-}\widetilde{\left| 
\mathcal{E}^{\mathrm{out}}\right\rangle }  \label{Identify}
\end{eqnarray}
It relates the transfer and scattering amplitudes.
We decompose the scattering equations (\ref{Smatrix}) on the two
components and use (\ref{Identify}) to rewrite them as 
\begin{eqnarray}
\pi _{+}\left| \mathcal{E}_{\mathrm{R}}\right\rangle &=&\pi _{+}\widetilde{S}
\left( \pi _{+} \left| \mathcal{E}_{\mathrm{L}}\right\rangle +\pi _{-}\left| 
\mathcal{E}_{\mathrm{R}}\right\rangle \right)  \nonumber \\
\pi _{-}\left| \mathcal{E}_{\mathrm{L}}\right\rangle &=&\pi _{-}\widetilde{S}
\left( \pi _{+}\left| \mathcal{E}_{\mathrm{L}}\right\rangle +\pi _{-}\left| 
\mathcal{E}_{\mathrm{R}}\right\rangle \right)
\end{eqnarray}
This linear system may be put under a matrix form 
\begin{equation}
\left( \pi _{-}-\widetilde{S}\pi _{+}\right) \left| \mathcal{E}_{\mathrm{L}}
\right\rangle =-\left( \pi _{+}-\widetilde{S}\pi _{-}\right) \left| 
\mathcal{E}_{\mathrm{R}}\right\rangle
\end{equation}
It is equivalent to the transfer equation (\ref{Tmatrix}) 
with the $T-$matrix obtained as
\begin{equation}
T=-\left( \pi _{-}-\widetilde{S} \pi_{+}\right) ^{-1}
\left( \pi _{+}-\widetilde{S}\pi _{-}\right)
\label{StoT}
\end{equation}

The converse transformation is obtained by performing the same manipulations in the 
reverse order. Starting from the transfer equation (\ref{Tmatrix}) and using 
(\ref{Identify}), one obtains a linear system which is equivalent 
to the scattering equation (\ref{SmatrixTilde}) with 
\begin{equation}
\widetilde{S}
=-\left( \pi _{-}-T\pi _{+}\right) ^{-1}\left( \pi _{+}-T\pi _{-}\right)
\label{TtoS}
\end{equation}
The relations (\ref{StoT}) and (\ref{TtoS}) have the same form. 
They represent an idempotent homographic transformation in
the space $\mathcal{M}_{2}\left( \mathbb{C}\right)$,
care being taken for the non-commutativity of multiplications in this space.
When inverting algebraically the homographic relations 
(\ref{StoT}) and (\ref{TtoS}), one obtains equivalent expressions 
\begin{eqnarray}
&&\widetilde{S}=\left( \pi _{+}+\pi _{-}T\right)
\left( \pi _{-}+\pi _{+}T\right) ^{-1} \nonumber \\
&&T=\left( \pi _{+}+\pi _{-}\widetilde{S}\right)
\left( \pi _{-}+\pi _{+}\widetilde{S}\right) ^{-1}
\end{eqnarray}
Other equivalent expressions are obtained from the equalities 
\begin{eqnarray}
\left( \pi _{-}-\widetilde{S}\pi _{+}\right) 
\left( \pi _{-}-T\pi_{+}\right) &=& I \nonumber \\
\left( \pi _{-}+\pi _{+}\widetilde{S}\right) 
\left( \pi _{-}+\pi_{+}T\right) &=& I
\label{SandTsym}
\end{eqnarray}

All these expressions may be written in terms of the scattering 
and transfer amplitudes 
\begin{eqnarray}
a=\frac{1}{t} &&\qquad b=-\frac{\overline{r}}{t} \nonumber \\ 
c=\frac{r}{t} &&\qquad d=\frac{t\overline{t}-r\overline{r}}{t} \nonumber \\
r=\frac{c}{a} &&\qquad \overline{t}=\frac{ad-bc}{a} \nonumber \\ 
t=\frac{1}{a} &&\qquad \overline{r}=-\frac{b}{a}
\label{SandTamplitudes}
\end{eqnarray}
The more formal homographic transformations written above are nevertheless
useful, as it will become clear in forthcoming calculations.

\subsection{Composition of optical networks}

\begin{figure}[tbh]
\centerline{\psfig{figure=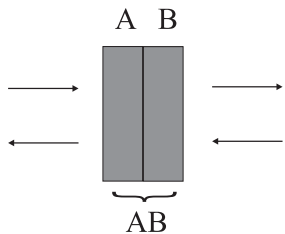,height=3cm}}
\caption{\label{FigComposition}Composition of networks~:
two networks labelled A and B are piled up to build
up a network AB.}
\end{figure}

The $T-$matrices are perfectly adapted to the composition of
optical networks corresponding to a piling up process (see Figure 
\ref{FigComposition}).

\vspace*{4mm}\vspace*{4mm}

On each network, the transfer equations are written as
\begin{eqnarray}
\left| \mathcal{E}_\mathrm{L} \net{A} \right\rangle &=& T \net{A} \left| \mathcal{E}
_{\mathrm{R}} \net{A} \right\rangle \nonumber\\ 
\left| \mathcal{E}_{\mathrm{L}}
\net{B} \right\rangle &=& T \net{B} \left| \mathcal{E}_{\mathrm{R}} \net{B}
\right\rangle
\end{eqnarray}
The brackets $\net{\ }$ specify the network for which the $T-$matrix or
field column is written. 
Identifying the fields according to Figure (\ref{FigComposition}) 
\begin{eqnarray}
&&\left| \mathcal{E}_{\mathrm{L}} \net{AB} \right\rangle \equiv
\left| \mathcal{E}_{\mathrm{L}} \net{A} \right\rangle \qquad
\left| \mathcal{E}_{\mathrm{R}} \net{A} \right\rangle \equiv
\left| \mathcal{E}_{\mathrm{L}} \net{B} \right\rangle \nonumber \\
&&\left| \mathcal{E}_{\mathrm{R}} \net{AB} \right\rangle \equiv 
\left| \mathcal{E}_{\mathrm{R}} \net{B} \right\rangle
\label{identifyCompo}
\end{eqnarray}
we deduce that the piling up process is equivalent to the product 
of $T$-matrices 
\begin{eqnarray}
\left| \mathcal{E}_{\mathrm{L}} \net{AB} \right\rangle &=&
T \net{AB}  \left| \mathcal{E}_{\mathrm{R}} \net{AB} \right\rangle
\nonumber \\ 
T \net{AB}  &=& T \net{A}  T \net{B}  
\label{compoT}
\end{eqnarray}
We have assumed the two networks to be in the immediate vicinity of each other
but without any electronic exchange between them, which again corresponds to the
assumption of thick enough layers.

\subsection{Elementary networks}

We now study two elementary networks, that is the traversal of an 
interface and the propagation over a given length inside a dielectric medium. 

For the scattering at the plane interface between two media with indices 
$n_{0}$ and $n_{1}$, we write the reflection and transmission amplitudes 
as the Fresnel scattering amplitudes \cite{LLfresnel}. 
Reflection amplitudes $r^p \net{Int}$ are obtained from characteristic 
impedances $z^p$ defined for plane waves with polarization $p$
in each medium and from the continuity equations at the interface
\begin{eqnarray}
r^p \net{Int} &&= - \overline{r}^p \net{Int} = \frac{1-z^p}{1+z^p} \nonumber \\
z^\mathrm{TE} &&= \frac{n_1 \cos\theta _1} {n_0 \cos \theta_0} 
= \frac{\kappa_1} {\kappa_0} \nonumber \\ 
z^\mathrm{TM} &&= \frac{n_1 \cos\theta _0} {n_0 \cos \theta_1} 
= \frac{\varepsilon _{1}\kappa_{0}}{\varepsilon_0  \kappa _{1}}
\label{r01}
\end{eqnarray}
Then the transmission amplitudes are obtained as 
\begin{equation}
\sqrt{\frac{\kappa_1}{\kappa_0}} t^p \net{Int}  
= \sqrt{\frac{\kappa_0}{\kappa_1}} \overline{t}^p  \net{Int}  
= \sqrt{1- \left( r^p \net{Int} \right) ^{2} }
\end{equation}
We deduce the expression of the transfer matrix 
\begin{eqnarray}
&&T^p \net{Int} = \sqrt{\frac{\kappa_1}{\kappa_0}}
\frac{1}{\sqrt{2 \sinh \beta^p }}
\left( 
\begin{tabular}{ll}
$e^{\frac{\beta^p}{2}}\quad $ & $-e^{-\frac{\beta^p }{2}}$ \\ 
$-e^{-\frac{\beta^p}{2}}$ & $\quad e^{\frac{\beta^p }{2}}$%
\end{tabular}
\right)  \nonumber \\
&&\beta ^p = \ln \frac{z^p + 1}{z^p - 1} 
\label{Tint}
\end{eqnarray}

We now consider the process of field propagation over a propagation length $\ell $ 
inside a dielectric medium characterized by a permittivity $\varepsilon$.
For this elementary network, the $T-$matrix has the simple form 
\begin{eqnarray}
&&T \net{Prop} =
\left( 
\begin{tabular}{ll}
$e^{\alpha}$ & $0$ \\ 
$0$ & $e^{-\alpha}$%
\end{tabular}
\right) \nonumber \\
&&\alpha = \kappa \ell = \sqrt{\varepsilon 
\frac{\xi ^2}{c^2}+\mathbf{k}^2}  \ \ell 
\label{Tprop}
\end{eqnarray}
The optical depth $\alpha$ does not depend on the polarization. 

Note that the composition is commutative within the class of interfaces 
or that of propagations~: it corresponds to the multiplication of the $z-$parameters 
for interfaces and to the addition of $\alpha -$parameters for propagations. But the 
composition is no longer commutative when interfaces and propagations are piled up.

\subsection{Reciprocity theorem}

We now prove a reciprocity theorem obeyed by arbitrary dielectric multilayers,
{\it i.e.} networks obtained by piling up interfaces and propagations.

To this aim, we first remark that the ratio of the two transmission 
amplitudes is related to the determinant of the $T-$matrix 
\begin{equation}
\frac{\overline{t}}{t}=ad-bc=\det T  
\label{determinant0}
\end{equation}
This follows from the relations (\ref{SandTamplitudes}) between $S-$ and $T-$amplitudes 
for an arbitrary network. 
Then, it is clear from (\ref{compoT}) that the determinant of 
$T$ is simply multiplied under composition 
\begin{equation}
\det T \net{AB} = \det T \net{A} \ \det T \net{B} 
\end{equation}
For the two kinds of elementary networks studied previously (see eqs \ref{Tint}-\ref{Tprop}), 
the determinant of $T$ is the ratio of the values of $\kappa$ at the right and left sides 
of the network
\begin{equation}
\det T = \frac{\kappa _\mathrm{R}}{\kappa _\mathrm{L}}
\label{RecipTheorem}
\end{equation}
It follows that this relation is valid for any optical network composed
by piling up interfaces and propagations.

In the particular case where the network has its two ports corresponding to vacuum,
which is the case for a mirror, the values of $\kappa$ are equal on its 
two sides and the $T-$matrix has a unit determinant 
\begin{equation}
\det T = 1 \qquad \overline{t}=t
\label{RecipTheorem1}
\end{equation}
Note that reciprocity corresponds to a symmetrical $S-$matrix 
and has to be distinguished from the spatial symmetry of the network with respect to its 
mediane plane which entails $\overline{r}=r$. 

This theorem is the specific form, when the symmetry of plane mirrors is assumed, 
of the general reciprocity theorem demonstrated by Casimir \cite{Casimir45}
as an extension to electromagnetism of Onsager's microreversibility theorem
\cite{Onsager31}. We have disregarded any static magnetic field which could affect 
these reciprocity relations.

\subsection{Slabs and multilayers}

We now consider the dielectric slabs and multilayers as composed networks and we 
deduce their transfer and scattering amplitudes from the preceding results. 

The slab is obtained by piling up a vacuum/matter interface with indices $n_0=1$ and 
$n_1$ at its left and righthand sides, propagation over a length $\ell$ inside 
matter, and a matter/vacuum interface with now $n_1$ and $n_0=1$ at its 
left and righthand sides. We denote $T\net{Int}$ the $T-$matrix associated 
with the first interface and obtain the $T-$matrix associated with the second 
interface as the inverse of $T\net{Int}$.
As a consequence of the composition law (\ref{compoT}), the $T-$matrix associated
with the slab is obtained as 
\begin{equation}
T \net{Slab} = T \net{Int} T \net{Prop} T \net{Int} ^{-1}
\end{equation}

Using the expressions (\ref{Tint},\ref{Tprop}) of $T \net{Int}$
and $T \net{Prop}$, we evaluate $T \net{Slab}$ as 
\begin{equation}
T\net{Slab} = \frac{1}{\sinh \beta}
\left( 
\begin{tabular}{cc}
$\sinh \left( \beta +\alpha \right)$ & $\sinh \alpha $ \\ 
$-\sinh \alpha $ & $\sinh \left( \beta -\alpha\right) $%
\end{tabular}
\right)  
\label{Tslab}
\end{equation}
We deduce the form of the $S-$matrix which is simultaneously reciprocal 
($\overline{t}=t$) and symmetrical in the exchange of its two ports 
($\overline{r}=r$) 
\begin{equation}
S\net{Slab} = \frac{1}{\sinh \left( \beta +\alpha \right)}
\left( 
\begin{tabular}{cc}
$-\sinh \alpha $ & $\sinh \beta $ \\ 
$\sinh \beta $ & $-\sinh \alpha $%
\end{tabular}
\right)
\label{slab}
\end{equation}
In the limiting case of a small thickness $\alpha \rightarrow 0$, we find 
$t \net{Slab} \rightarrow 1$ and $r \net{Slab} \rightarrow 0$, which
means that the slab tends to become transparent. In this case indeed, 
the propagation can be forgotten and the two inverse interfaces have 
their effects cancelled by each other.

The opposite limiting case of a large thickness is often considered since it fits
the usual experimental situations. More precisely, experiments are performed 
with metallic mirrors having a thickness much larger than the plasma wavelength.
This is why the limit of a total extinction of the field through 
the medium is assumed in most calculations. This corresponds to the so-called 
`bulk limit' with $e^{-\alpha }\rightarrow 0$ and 
$r \net{Slab} \rightarrow -e^{-\beta }= r \net{Int}$ in eq.(\ref{slab})~: 
the reflection amplitude is determined entirely by the first interface. 
Let us emphasize however that the bulk limit raises several delicate problems.
First, the transmission amplitude $t\net{Slab}$ vanishes in this limit 
so that the $T-$matrix is not defined, with the drawback of 
invalidating the general method used in the present paper.
Then, the bulk limit cannot be met in the case of non absorbing media where 
$e^{-\alpha }$ remains a complex number with unit modulus for any value of $\ell$. 
Even in the presence of absorption, a large value of the width $\ell$ does not 
necessarily imply a large value of the optical thickness $\alpha$ since $\kappa$ 
may go to zero at normal incidence and zero frequency, leading to a transparent 
slab in contrast with the results of the bulk limit. 
Therefore a reliable calculation must consider
the experimental situation of mirrors with a large but finite thickness. 
In the present paper, we consider the general case of arbitrary mirrors
and test the reliability of the bulk limit in the end of the calculations.

We can deal with the case of dielectric multilayers similarly. If we consider as
an example the multilayer obtained by piling up a vacuum/matter interface with 
indices $n_0=1$ and $n_1$ at its left and righthand sides, propagation over a 
length $\ell_1$ inside the medium 1, an interface between media 1 and 2, 
propagation over a length $\ell_2$ inside the medium 2, and an interface 
between medium 2 and vacuum, its $T-$matrix is obtained as the product 
\begin{eqnarray}
T \net{Multilayer} &=& T \net{Int01} T \net{Prop1} T \net{Int12} \nonumber \\
&& \times T \net{Prop2} T \net{Int20} 
\end{eqnarray}
Alternatively, the same multilayer may be obtained by piling up two 
slabs each corresponding to one of the layers
\begin{eqnarray}
T \net{Multilayer} &=& T \net{Slab010} T \net{Slab020} 
\end{eqnarray}
In the last two equations, the indices specify the different interfaces,
propagations or slabs using an obvious convention. 

Since any multilayer mirror is obtained by piling up slabs connecting two vacuum ports
and thus obeying the reciprocity relation $\overline{t}=t$, we can use a simple form 
of the composition law written in terms of scattering amplitudes \cite{Lambrecht97} 
\begin{eqnarray}
r\netsimple{AB} &=&r\netsimple{A}+\frac{t\netsimple{A}^{2} r\netsimple{B}}
{1-\overline{r}\netsimple{A} r\netsimple{B}}  \qquad
\overline{r}\netsimple{AB} =\overline{r}\netsimple{B}
+\frac{\overline{r}\netsimple{A} t\netsimple{B}^{2}}
{1-\overline{r}\netsimple{A} r\netsimple{B}} \nonumber \\
t\netsimple{AB} &=&\frac{t\netsimple{A} t\netsimple{B}}
{1-\overline{r}\netsimple{A} r\netsimple{B}}  
\label{compoSlab}
\end{eqnarray}
For readibility, we have specified the networks by using subscripts rather than brackets.
We will proceed similarly in forthcoming specific computations. Iterating this composition 
law, we can compute the scattering amplitudes for any dielectric multilayer.
This systematic technique is quite similar to the classical computation techniques used 
for studying multilayers \cite{Abeles55}. It is generalized to the full quantum
treatment in the next section. It also leads in the following to general results 
constraining the variation of the Casimir force for arbitrary dielectric mirrors.
It reproduces the known results for the multilayer systems which have already 
been studied \cite{Zhou95,Bordag01}.

\section{Quantum treatment of lossy mirrors}

Up to now, we have performed a classical analysis which is not 
sufficient for the purpose of describing the scattering of vacuum fluctuations. 
Real mirrors consist of absorbing media which scatter incident fields to 
spontaneous emission modes and reciprocally scatter fluctuations from noise modes 
to the modes of interest. The $S-$matrix calculated previously cannot be unitary 
for a lossy mirror but it should be the restriction to the modes of interest 
of a larger $S-$matrix which includes the noise modes and obeys unitarity. 
In the present section, we characterize the additional fluctuations for a lossy mirror
by using the corresponding `optical theorem', that is also the unitarity of the 
larger $S-$matrix (see \cite{Barnett96,Courty00} and references therein).

We assume that the scattering restricted to the modes of interest still fulfills the 
symmetry of plane mirrors considered in the previous classical calculations. 
This amounts to neglect multiple scattering processes which could couple different modes
through their coupling with noise modes. Except for this assumption, we consider arbitrary 
dissipative media and discuss the optical theorem in the scattering and transfer points of 
view. We use the latter one to deal with composition of additional fluctuations 
when lossy mirrors are piled up.

\subsection{Noise in the scattering approach}

Should we use the previous classical equations for the quantum amplitudes, 
we would find that the output fields cannot obey the canonical commutators, 
except in the particular case of lossless mirrors. 
This implies that the input/output transformation for quantum field must
include additional fluctuations superimposed to the classical equations
\begin{equation}
\left| e^\mathrm{out} \right\rangle = S \left| e^\mathrm{in}\right\rangle 
+ \left| F\right\rangle 
\label{SmatrixNoise}
\end{equation}
$\left| e^\mathrm{out} \right\rangle$ and $\left| e^\mathrm{in} \right\rangle$ 
are defined as in (\ref{Smatrix}) with the quantum amplitudes $e$ in place of 
the classical fields $\mathcal{E}$, $S$ is the same matrix as previously and
$\left| F\right\rangle$ is a twofold column matrix describing the additional
fluctuations. All these quantities depend on the quantum number $m$ which
is common to all fields coupled in the scattering process.

\vspace*{4mm}\begin{figure}[tbh]
\centerline{\psfig{figure=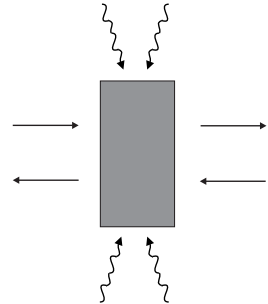,height=4cm}}
\caption{Representation of a dissipative network, with
additional fluctuations coming from the noise modes.}
\label{FigAddFluct}
\end{figure}\vspace*{4mm}

The additional fluctuations are linear superpositions of all modes coupled to 
the main modes $e_{m}^{\phi }$ by the microscopic couplings which cause 
absorption. As an example, the atoms constituting a dielectric medium couple 
the main modes to all electromagnetic modes through spontaneous emission 
processes, represented symbolically by the wavy arrows on 
Figure \ref{FigAddFluct}. 
The stationarity assumption implies that only modes having the
same frequencies are coupled. In particular, it forbids 
parametric couplings which could couple modes with different frequencies and
`squeeze' the vacuum fluctuations \cite{Reynaud90}. The whole scattering
matrix which takes into account all coupled field modes is unitary and
this basic property makes the canonical commutation relations compatible for
input and output fields. In contrast, the reduced scattering matrix containing 
only the classical scattering amplitudes coupling the main modes 
$e_{m}^{\phi }$ is not unitary, except in the particular case of lossless mirrors.

In order to write the unitarity property of the whole scattering matrix, 
it is convenient to represent the additional fluctuations $\left| F\right\rangle$ 
by introducing auxiliary noise modes $\left| f \right\rangle$ 
and auxiliary noise amplitudes gathered in a noise matrix $S^\prime$ 
\begin{equation}
\left| F\right\rangle =S^\prime \left| f\right\rangle \qquad 
S^\prime =\left( 
\begin{tabular}{ll}
$r^\prime $ & $t^\prime $ \\ 
$t^\prime $ & $\overline{r}^\prime $%
\end{tabular}
\right)
\end{equation}
The components of the twofold column $\left| f\right\rangle$ are 
defined to have the same canonical commutators as the input fields in the main 
modes. In fact, they are linear superpositions of the input vacuum modes 
responsible for the fluctuation process. They are defined up to an 
ambiguity~: any canonical transformation of the noise modes leads to an equivalent
representation of the additional fluctuations, which corresponds to a different form 
for the noise amplitudes while leading to the same physical results at the end of the 
computations.

For any of these equivalent representations, the norm matrix $S^{\prime
}S^{\prime \ \dagger }$ has the same expression determined by the optical
theorem, that is the unitarity condition for the whole scattering process, 
\begin{equation}
SS^\dagger +S^\prime S^{\prime \ \dagger }=I  \label{unitarity}
\end{equation}
where $I$ is the $2 \times 2$ unity matrix. 
This is easily proven by a direct inspection of the explicit expressions of the
commutators of the output fields. The same inspection shows that noise
modes corresponding to different values of $m$ are not correlated to each other. 
Condition (\ref{unitarity}) is made more explicit when $SS^\dagger$ and 
$S^\prime S^{\prime \ \dagger}$ are developed in terms of scattering amplitudes 
\begin{eqnarray}
rr^{\ast }+tt^{\ast }+r^\prime r^{\prime \ast }+t^\prime t^{\prime \ast
} &=&tt^{\ast }+\overline{r}\overline{r}^{\ast }+t^\prime t^{\prime \ast }+
\overline{r}^\prime \overline{r}^{\prime \ast }  \nonumber \\
&=&1  \nonumber \\
rt^{\ast }+t\overline{r}^{\ast }+r^\prime t^{\prime \ast }+t^\prime 
\overline{r}^{\prime \ast } &=&tr^{\ast }+\overline{r}t^{\ast }+t^{\prime
}r^{\prime \ast }+\overline{r}^\prime t^{\prime \ast } \nonumber \\
&=&0
\end{eqnarray}
More detailed discussions are presented for the case of the slab 
in appendix \ref{AppSlab}.

The description of noise may as well be represented with the alternative 
representation (\ref{SmatrixTilde}) of the scattering process
\begin{eqnarray}
&&\widetilde{\left| e^{\mathrm{out}} \right\rangle } = 
\widetilde{S}\left| e^{\mathrm{in}}\right\rangle 
+\widetilde{\left| F\right\rangle } \nonumber \\
&&\widetilde{\left| e^{\mathrm{out}} \right\rangle } = 
\eta \left| e^{\mathrm{out}} \right\rangle \qquad
\widetilde{\left| F\right\rangle } = \eta \left| F\right\rangle
\label{SmatrixTildeNoise}
\end{eqnarray}
The additional fluctuations are then represented in terms of the same 
noise modes and of a modified noise matrix 
\begin{eqnarray}
\widetilde{\left| F\right\rangle } &=& 
\widetilde{S^\prime} \left| f \right\rangle \qquad 
\widetilde{S^\prime} = \eta S^\prime  \nonumber \\
&&\widetilde{S^\prime} \widetilde{S^\prime} ^\dagger = 
I - \widetilde{S}\widetilde{S}^\dagger   
\label{unitarityTilde}
\end{eqnarray}

\subsection{Noise in the transfer approach}

We now present the description of additional fluctuations in the transfer approach.
Performing the same manipulations as in the previous section, we transform 
equation (\ref{SmatrixTildeNoise}) into 
\begin{equation}
\left( \pi _{-}-\widetilde{S}\pi _{+} \right) \left| e_{\mathrm{L}}
\right\rangle = -\left( \pi _{+}-\widetilde{S}\pi _{-}\right) \left| 
e_{\mathrm{R}}\right\rangle +\widetilde{\left| F\right\rangle }
\end{equation}
We thus get transfer equations with additional fluctuations described by
a twofold column $\left| G\right\rangle$
\begin{eqnarray}
&&\left| e_{\mathrm{L}}\right\rangle = T \left| e_{\mathrm{R}}\right\rangle 
+ \left| G\right\rangle \nonumber \\
&&\left| G\right\rangle = \left(\pi _{-}-\widetilde{S}\pi _{+}\right) ^{-1}
\widetilde{\left| F\right\rangle }  
\label{FtoG}
\end{eqnarray}
The $T-$ matrix has the same expression (\ref{StoT}) as previously and the 
additional fluctuations $\left| G\right\rangle $
are a linear expression of the fluctuations $\left| F\right\rangle $
defined in the scattering approach. This linear relation may be written 
under alternative forms by using the relations (\ref{SandTsym})
\begin{eqnarray}
\left| F\right\rangle &&= \left(\pi _{-}-\widetilde{S}\pi _{+}\right) 
\widetilde{\left| G\right\rangle }  
= \left( \pi _{-}-T\pi_{+}\right) ^{-1}
\widetilde{\left| G\right\rangle }  \nonumber \\
\left| G\right\rangle &&= \left( \pi _{-}-T\pi_{+}\right) 
\widetilde{\left| F\right\rangle }  
\end{eqnarray}
 
In the scattering approach, the norm of additional fluctuations is
described by matrices $S^\prime S^{\prime \ \dagger}$ and
$\widetilde{S^\prime} \widetilde{S^\prime} ^\dagger$ which are themselves
determined by the optical theorem (\ref{unitarity}) or (\ref{unitarityTilde}). 
In order to translate these properties to the transfer approach,
we rewrite (\ref{FtoG}) in terms of the canonical noise modes 
$\left| f\right\rangle$ and of noise amplitudes gathered in a matrix $T^\prime $ 
\begin{eqnarray}
&&\left| G\right\rangle =T^\prime \left| f\right\rangle \nonumber \\
T^\prime &&=\left( \pi _{-} - \widetilde{S}\pi _{+}\right) ^{-1}
\widetilde{S^\prime }=\left( \pi _{-}-T\pi _{+}\right) \widetilde{S^\prime }
\end{eqnarray}
The associated norm matrix is
\begin{eqnarray}
T^\prime T^{\prime \ \dagger}&&=\left( \pi _{-}-T\pi _{+}\right) 
\widetilde{S^\prime } \widetilde{S^\prime } ^\dagger 
\left( \pi _{-}-T\pi _{+}\right) ^\dagger 
\end{eqnarray}
Using equations (\ref{unitarityTilde}) and (\ref{TtoS}), we rewrite it as 
\begin{equation}
T^\prime T^{\prime \ \dagger }=T\Phi T^\dagger - \Phi \qquad 
\Phi =\pi _{+}-\pi _{-}
\label{unitaryT}
\end{equation}
$\Phi $ is a diagonal matrix with two eigenvalues representing the directions 
of propagation $\phi=\pm 1$ of the field.

\subsection{Composition of dissipative networks}

Using these tools, we now write composition laws for the
fluctuations and their norms.

We start from transfer equations written for each network A and B 
\begin{eqnarray}
\left| e_{\mathrm{L}} \net{A} \right\rangle &=& T \net{A} \left| e_{\mathrm{R}}
\net{A} \right\rangle +\left| G \net{A} \right\rangle \nonumber \\
\left| e_{\mathrm{L}} \net{B} \right\rangle &=& T \net{B} \left| e_{\mathrm{R}} 
\net{B} \right\rangle +\left| G \net{B} \right\rangle
\end{eqnarray}
Using the identifications (\ref{identifyCompo}) associated with the composition
law, we deduce for the composed network 
\begin{eqnarray}
\left| e_{\mathrm{L}} \net{AB} \right\rangle &=&T \net{AB} \left| e_{\mathrm{R}}
\net{AB} \right\rangle +\left| G \net{AB} \right\rangle \nonumber \\
\left| G \net{AB} \right\rangle &=&
\left| G \net{A} \right\rangle +T \net{A} \left| G \net{B}\right\rangle  
\label{compoG}
\end{eqnarray}
The fluctuations $\left| G \net{AB} \right\rangle$ are a linear superposition 
of fluctuations $\left| G \net{A} \right\rangle$ and 
$\left| G \net{B} \right\rangle$ added in A and B.

In order to obtain the composition law for the norm matrices, we 
develop the additional fluctuations $\left| G \net{AB} \right\rangle$ 
on the canonical noise modes associated with the two elements 
\begin{equation}
\left| G \net{AB} \right\rangle = T^\prime \net{A} \left| f \net{A}
\right\rangle + T \net{A} T^\prime \net{B} \left| f\net{B} \right\rangle
\end{equation}
Since the noise modes associated with different elements are uncorrelated,
$\left| G \net{AB} \right\rangle$ may be rewritten in terms of
new canonical noise modes and new noise amplitudes such that
\begin{eqnarray}
\left| G \net{AB} \right\rangle &=& T^\prime\net{AB} 
\left| f \net{AB} \right\rangle \nonumber \\
T^\prime\net{AB} T^\prime \net{AB} ^\dagger &=&T^\prime\net{A} 
T^\prime \net{A} ^\dagger \nonumber \\
&+&T\net{A} T^\prime\net{B} 
T^\prime \net{B} ^\dagger T\net{A} ^\dagger 
\label{compoNoiseT}
\end{eqnarray}
Using expression (\ref{unitaryT}) of the optical theorem for both networks
A and B, we deduce that the composed network AB obeys the same relation
\begin{eqnarray}
T^\prime\net{AB} T^\prime \net{AB} ^\dagger &&=
T \net{A} \Phi T\net{A} ^\dagger - \Phi \nonumber \\
&&+ T \net{A} \left( T \net{B} \Phi T\net{B} ^\dagger - \Phi \right) 
T\net{A} ^\dagger \nonumber \\
&&= T \net{AB} \Phi T \net{AB} ^\dagger - \Phi
\end{eqnarray}
Equivalently, the $S-$matrix of the composed network AB obeys the optical theorem 
(\ref{unitarity}) as soon as the two networks A and B do.

\subsection{Resonance for cavity fields}

We have studied the scattering or, equivalently, the lefthand/righthand transfer of 
fields by a composed network AB. We want now to characterize the properties of the 
fields inside the cavity formed between A and B. This problem will play a key role in 
the evaluation of the Casimir force (see next section).

\vspace*{4mm}\begin{figure}[tbh]
\centerline{\psfig{figure=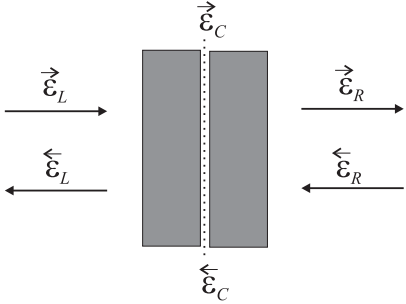,height=4cm}}
\caption{Cavity formed within a composed network~: L and R denote the fields
at left and right sides of the network whereas C denotes the cavity fields.}
\label{FigCavity}
\end{figure}\vspace*{4mm}

The situation is illustrated by Figure \ref{FigCavity} which, in contrast
to Figure \ref{FigComposition}, keeps the trace of the intracavity fields.
In algebraic terms, the cavity fields are defined by rewriting 
the identifications (\ref{identifyCompo}) as 
\begin{eqnarray}
&&\left| e_{\mathrm{L}} \net{AB} \right\rangle \equiv
\left| e_{\mathrm{L}} \net{A} \right\rangle \qquad
\left| e_{\mathrm{R}} \net{AB} \right\rangle \equiv 
\left| e_{\mathrm{R}} \net{B} \right\rangle \nonumber \\
&&\left| e_{\mathrm{C}} \net{AB} \right\rangle \equiv
\left| e_{\mathrm{R}} \net{A} \right\rangle =
\left| e_{\mathrm{L}} \net{B} \right\rangle 
\end{eqnarray}
From now on, we drop the label $\net{AB}$ for the composed network 
and use subscripts for the networks A and B. 

In order to express the cavity fields in terms of the input modes and additional 
fluctuations, we first write the cavity fields $\left| e_\mathrm{C} \right \rangle$ 
in terms of the righthand ones $\left| e_\mathrm{R} \right \rangle$ 
\begin{equation}
\left| e_{\mathrm{C}} \right \rangle =T \netsimple{B} \left| e_{\mathrm{R}} \right\rangle
+\left| G\netsimple{B} \right\rangle 
\end{equation}
We then identify the two components of $\left| e_\mathrm{R} \right \rangle$ as
\begin{eqnarray}
\pi _{+}\left| e_{\mathrm{R}}\right\rangle &=& \pi _{+}\widetilde{\left| 
e^{\mathrm{out}}\right\rangle }=\pi _{+}\left( \widetilde{S}\left| e^{\mathrm{in}}
\right\rangle +\widetilde{ \left| F\right\rangle }\right) \nonumber \\
\pi_{-}\left| e_{\mathrm{R}}\right\rangle &=& \pi _{-}\left| e^{\mathrm{in}}\right\rangle
\end{eqnarray}
Using the expression of $\widetilde{ \left| F\right\rangle}$ in terms of 
$\left| G\right\rangle$ and the composition law (\ref{compoG}) for 
$\left| G\right\rangle$, we deduce 
\begin{eqnarray}
&&\left| e_{\mathrm{C}}\right\rangle = 
R \left| e^{\mathrm{in}}\right\rangle + R^\prime \netsimple{A} \left| f\netsimple{A}
\right\rangle + R^\prime \netsimple{B} \left| f \netsimple{B} \right\rangle \nonumber \\
&&R = T\netsimple{B} N \qquad 
N= \left( \pi _{+}\widetilde{S}+\pi _{-}\right) =
\left( \pi _{-}+\pi _{+}T\right) ^{-1} \nonumber \\
&&R^\prime \netsimple{A} = T\netsimple{B} P T^\prime \netsimple{A} \qquad 
P= - N \pi _{+} \nonumber \\
&&R^\prime \netsimple{B} = \left( I + T\netsimple{B} P T\netsimple{A} \right) 
T^\prime \netsimple{B}   
\end{eqnarray}

As already explained, the unitarity of scattering entails that the output
fields have the same commutators as the input ones. But this is not the case for
the cavity fields which have their commutators determined by the matrix 
\begin{equation}
\mathcal{G} = RR^\dagger +R^\prime \netsimple{A} R^{\prime\ \dagger} \netsimple{A} 
+R^\prime \netsimple{B} R^{\prime\ \dagger} \netsimple{B} 
\end{equation}
Expanding this quadratic form and using the composition law (\ref{compoNoiseT}),
we rewrite $\mathcal{G}$ as 
\begin{eqnarray}
\mathcal{G} &=& T\netsimple{B} NN^\dagger T\netsimple{B}^\dagger +T\netsimple{B}
PT^\prime T^{\prime\ \dagger } P^\dagger T\netsimple{B}^\dagger \nonumber \\
&&+ T\netsimple{B} P T\netsimple{A} T^\prime \netsimple{B} T ^{\prime\ \dagger} 
\netsimple{B} \nonumber \\
&&+ T^\prime \netsimple{B} T^{\prime\ \dagger} \netsimple{B} T\netsimple{A}^\dagger
P^\dagger T\netsimple{B}^\dagger \nonumber \\
&&+ T^\prime \netsimple{B} T^{\prime\ \dagger} \netsimple{B} 
\end{eqnarray}
Using relation (\ref{unitaryT}) for the three networks A, B and AB,
we obtain a simpler expression after a few rearrangements
\begin{equation}
\mathcal{G} = - \Phi - T\netsimple{B} P T\netsimple{A} \Phi  
-\Phi T\netsimple{A}^\dagger P^\dagger T\netsimple{B}^\dagger 
\end{equation}

We now proceed to explicit calculations of these matrices. We note that 
$P = - t \pi_+$ where $t$ is the transmission amplitude of the network AB and deduce
\begin{equation}
-T\netsimple{B} P T\netsimple{A}\Phi =
t \left( 
\begin{tabular}{ll}
$a\netsimple{B} a\netsimple{A}$ & $-a\netsimple{B} b\netsimple{A}$ \\ 
$c\netsimple{B} a\netsimple{A}$ & $-c\netsimple{B} b\netsimple{A}$%
\end{tabular}
\right)
\end{equation}
$t$ is simply the inverse of the transfer amplitude $a$ associated with the network AB
(see eq.\ref{SandTamplitudes}) and the latter is deduced from the composition law 
(\ref{compoT})
\begin{equation}
t=\frac 1{a} \qquad a =a\netsimple{A} a\netsimple{B}
+b\netsimple{A} c\netsimple{B}
\end{equation}
Then, the transfer amplitudes of the networks A and B may be substituted 
by the associated scattering amplitudes, leading to 
\begin{equation}
-T\netsimple{B} P T\netsimple{A}\Phi =\frac{1}{1-\overline{r}\netsimple{A} r\netsimple{B}}
\left( 
\begin{tabular}{cc}
$1$ & $\overline{r}\netsimple{A}$ \\ 
$r\netsimple{B}$ & $\overline{r}\netsimple{A} r\netsimple{B}$%
\end{tabular}
\right)
\end{equation}
Collecting these results and proceeding to slight rearrangements, we finally get 
\begin{eqnarray}
\mathcal{G} &=&I+\frac{1}{1-\overline{r}\netsimple{A} r\netsimple{B}} \left( 
\begin{tabular}{cc}
$\overline{r}\netsimple{A} r\netsimple{B}$ & $\overline{r}\netsimple{A}$ \\ 
$r\netsimple{B}$ & $\overline{r}\netsimple{A} r\netsimple{B}$%
\end{tabular}
\right) \nonumber \\
&&+\frac{1}{\left( 1-\overline{r}\netsimple{A} r\netsimple{B}\right) ^\ast }
\left( 
\begin{tabular}{cc}
$\overline{r}\netsimple{A} r\netsimple{B}$ & $\overline{r}\netsimple{A}$ \\ 
$r\netsimple{B}$ & $\overline{r}\netsimple{A} r\netsimple{B}$%
\end{tabular}
\right) ^\dagger 
\end{eqnarray}
In the following we will use the diagonal terms of the matrix $\mathcal{G}$
to evaluate the Casimir force.

\subsection{Scattering on a Fabry-Perot cavity}

\begin{figure}[tbh]
\centerline{\psfig{figure=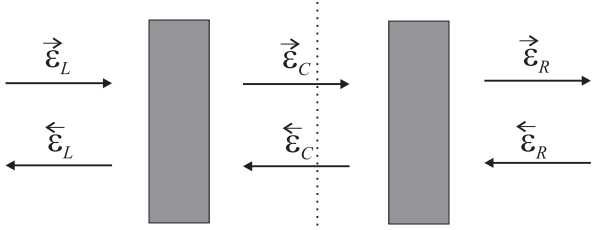,height=3cm}}
\caption{Representation of a Fabry-Perot cavity~: L and R denote the fields
at left and right sides of the cavity whereas C denotes the cavity fields inside
the Fabry-Perot cavity; these cavity fields are defined at an arbitrary position 
between the two mirrors.}
\label{FigFabryPerot}
\end{figure}

In order to prepare the evaluation of the Casimir force, we generalize the
preceding expression to the case of the Fabry-Perot cavity containing a zone of field 
propagation between the two mirrors M1 and M2 (see Figure \ref{FigFabryPerot}). 

\vspace*{4mm}\vspace*{4mm}

The distance between the two mirrors is denoted $L$ and the cavity fields 
are defined at an arbitrary position inside the cavity, say at distances 
$L_1$ from M1 and $L_2$ from M2 with $L_1 + L_2 = L$.
In these conditions, the study of the Fabry-Perot cavity is reduced to the 
problem studied in the preceding subsection through the following identifications~: 
the network A contains the mirror M1 and the propagation L1 with
$T \netsimple{A}  = T\netsimple{M1} T\netsimple{L1}$ while
the network B contains the propagation L2 and the mirror M2 with 
$T \netsimple{B} = T\netsimple{L2} T\netsimple{M2}$.
The transfer amplitudes for the networks A and B are derived from those
corresponding to M1 and M2 and from phase factors corresponding to
the propagations L1 and L2 
\begin{eqnarray}
t\netsimple{A} &=& \overline{t}\netsimple{A} = t_1 e^{-\alpha _1} \qquad 
\alpha_1 = \kappa_0 L_1 \nonumber \\
\overline{r}\netsimple{A}&=&\overline{r}_1 e^{-2\alpha _1} \qquad 
r\netsimple{A}=r_1 \nonumber \\
t\netsimple{B} &=& \overline{t}\netsimple{B} = e^{-\alpha _2} t_2 \qquad 
\alpha_2 = \kappa_0 L_2 \nonumber \\
\overline{r}\netsimple{B} &=& \overline{r}_2 \qquad r\netsimple{B} = r_2 e^{-2\alpha _2}
\end{eqnarray}
We have labeled the amplitudes for the mirrors M1 and M2 with mere indices 1 and 2; 
$\kappa_0$ is defined in vacuum. 
These results entail that the reflection amplitudes $\overline{r}\netsimple{A}$
and $r\netsimple{B}$ are seen from a point inside the cavity as the product of
phase factors by the reflection amplitudes $\overline{r}_1$ and $r_2$ seen from a 
point in the immediate vicinity of M1 and M2.

We then deduce the scattering amplitudes for the whole cavity 
\begin{eqnarray}
r &=& r_1 + \frac{t_1^2 r_2 e^{-2\alpha}}{D} \qquad
\overline{r} = \overline{r}_2 + \frac{\overline{r}_1 t_2^2 e^{-2\alpha}}{D} \nonumber \\
t &=& \overline{t} = \frac{t_1 t_2 e^{-\alpha}}{D} \nonumber \\
D &=& 1 - \overline{r}_1 r_2 e^{-2\alpha} \qquad \alpha =\alpha _1+\alpha _2  
\end{eqnarray}
and the expression of $\mathcal{G}$ 
\begin{eqnarray}
\mathcal{G} &=&I+\frac{1}{D}\left( 
\begin{tabular}{cc}
$\overline{r}_1 r_2 e^{-2\alpha}$ & $\overline{r}_1 e^{-2\alpha_1}$ \\ 
$r_2 e^{-2\alpha_2}$ & $\overline{r}_1 r_2 e^{-2\alpha}$ %
\end{tabular}
\right) \nonumber \\
&&+\frac{1}{D^\ast}\left( 
\begin{tabular}{cc}
$\overline{r}_1 r_2 e^{-2\alpha}$ & $\overline{r}_1 e^{-2\alpha_1}$ \\ 
$r_2 e^{-2\alpha_2}$ & $\overline{r}_1 r_2 e^{-2\alpha}$ %
\end{tabular}
\right) ^\dagger 
\label{Gmatrix}
\end{eqnarray}
The diagonal terms in the matrix $\mathcal{G}$ coincide with the Airy function 
\begin{eqnarray}
&&g=1+f+f^{\ast }=\frac{1-\left| \overline{r}_{1}r_{2}e^{-2\alpha }\right| ^{2}}
{\left| 1-\overline{r}_{1}r_{2}e^{-2\alpha }\right| ^{2}} \nonumber \\ 
&&f=\frac{\overline{r}_{1}r_{2}e^{-2\alpha }}{1-\overline{r}_{1}r_{2}e^{-2\alpha}}  
\label{gfunction}
\end{eqnarray}
This result will play the central role in the derivation of the Casimir
force in the next section. It means that the commutators of the intracavity
fields are not the same as those of the input or output fields. They
correspond to a spectral density modified through a multiplication
by the Airy function $g$. This is the basic property used 
in Cavity Quantum ElectroDynamics \cite{Haroche84}.

It is clear from the present derivation that this result has a
quite general status~: it is obtained for any inner field in any composed
network, assuming the symmetry of plane mirrors. 
This property was already known for non absorbing mirrors \cite{Jaekel91} 
and for lossy mirrors symmetrical with respect to their mediane plane \cite{Barnett98}. 
The present derivation proves that it is also valid for arbitrary dielectric multilayers
with dissipation. The final result only depends on the reflection amplitudes 
$\overline{r}_{1}$ and $r_{2}$ of the mirrors as they are seen from the inner side 
of the cavity. The reflection amplitudes seen from the outer side and the transmission
amplitudes do not appear in expressions (\ref{Gmatrix},\ref{gfunction}).
This can be interpreted as resulting from the unitarity of the 
whole scattering processes.

\section{Casimir force between real mirrors}

We may now deal with the radiation pressure of vacuum fields on the mirrors of a 
Fabry-Perot cavity. We show that the resulting Casimir force is a regular integral 
which can be written over real or imaginary frequencies. We then derive general 
constraints obeyed by the Casimir force for arbitrary dielectric mirrors.

\begin{widetext}

\subsection{Vacuum radiation pressure}

If we first consider a mirror isolated in vacuum, the radiation pressure is obtained 
by adding the contributions of the 4 fields coupled in the scattering process
\begin{eqnarray}
\left\langle P\right\rangle _{\mathrm{vac}} &=&\sum_{m}\ \hbar \omega _{m}\
\cos ^{2}\theta _{m}\ \left\langle e_{m\ \mathrm{L}}^{\rightarrow }\cdot e_{m\ 
\mathrm{L}}^{\rightarrow \ \dagger }+e_{m\ \mathrm{L}}^{\leftarrow }\cdot e_{m\ 
\mathrm{L}}^{\leftarrow \ \dagger }-e_{m\ \mathrm{R}}^{\rightarrow }\cdot e_{m\ 
\mathrm{R}}^{\rightarrow \ \dagger }-e_{m\ \mathrm{R}}^{\leftarrow }\cdot e_{m\ 
\mathrm{R}}^{\leftarrow \ \dagger }\right\rangle _{\mathrm{vac}}
\end{eqnarray}
The identification of these fields is given by Figure (\ref{FigNetwork}). 
We have developed the sum over $\phi $ and kept the symbol $m$ to represent the
quantum numbers $\left( \omega ,\mathbf{k},p\right)$. We assume that the whole 
system is in vacuum, that is at zero temperature, so that the anticommutators of 
input fields are given by relation (\ref{anticommVacuum}).
Since the commutators are the same for the output and input fields, 
the vacuum radiation pressure vanishes in the case of an isolated mirror.
In other words, the two sides of the mirror play equivalent roles so that no 
mean force can appear. 

When we consider two mirrors forming a Fabry-Perot cavity, the two sides of a given
mirror are no longer equivalent since one is an inner side and the other an outer side. 
It follows that the compensation observed for an isolated mirror does no longer hold, 
resulting in the appearance of the Casimir force.
In order to evaluate the force, we write the mean radiation pressures $\left\langle
P_{1}\right\rangle _{\mathrm{vac}}$ and $\left\langle P_{2}\right\rangle 
_{\mathrm{vac}}$ on mirrors M1 and M2 (see Figure \ref{FigFabryPerot}) 
\begin{eqnarray}
\left\langle P_{1}\right\rangle _{\mathrm{vac}} &=&\sum_{m}\ \hbar \omega _{m}\
\cos ^{2}\theta _{m}\ \left\langle e_{m\ \mathrm{L}}^{\rightarrow }\cdot e_{m\ 
\mathrm{L}}^{\rightarrow \ \dagger }+e_{m\ \mathrm{L}}^{\leftarrow }\cdot e_{m\ 
\mathrm{L}}^{\leftarrow \ \dagger }-e_{m\ \mathrm{C}}^{\rightarrow }\cdot e_{m\ 
\mathrm{C}}^{\rightarrow \ \dagger }-e_{m\ \mathrm{C}}^{\leftarrow }\cdot e_{m\ 
\mathrm{C}}^{\leftarrow \ \dagger }\right\rangle _{\mathrm{vac}}  \nonumber \\
\left\langle P_{2}\right\rangle _{\mathrm{vac}} &=&\sum_{m}\ \hbar \omega _{m}\
\cos ^{2}\theta _{m}\ \left\langle e_{m\ \mathrm{C}}^{\rightarrow }\cdot e_{m\ 
\mathrm{C}}^{\rightarrow \ \dagger }+e_{m\ \mathrm{C}}^{\leftarrow }\cdot e_{m\ 
\mathrm{C}}^{\leftarrow \ \dagger }-e_{m\ \mathrm{R}}^{\rightarrow }\cdot e_{m\ 
\mathrm{R}}^{\rightarrow \ \dagger }-e_{m\ \mathrm{R}}^{\leftarrow }\cdot e_{m\ 
\mathrm{R}}^{\leftarrow \ \dagger }\right\rangle _{\mathrm{vac}} 
\end{eqnarray}
\end{widetext}

For the same reasons as previously, the field anticommutators are given by 
(\ref{anticommVacuum}) for input and output fields. For intracavity fields, 
they are multiplied by the Airy function (\ref{gfunction}) like the commutators 
\begin{eqnarray}
\left\langle e_{m ^\prime \ \mathrm{C}}^{\phi ^\prime }\cdot e_{m\ \mathrm{C}}^{\phi
\ \dagger }\right\rangle _{\mathrm{vac}}
&=&\frac{1}{2}\left[ e_{m ^\prime \ \mathrm{C}}^{\phi ^\prime } , e_{m\ 
\mathrm{C}}^{\phi \ \dagger }\right] \nonumber \\
&=&\frac{1}{2}g_{m} \delta _{mm^\prime }\delta _{\phi \phi ^\prime} 
\label{anticommCav}
\end{eqnarray}
As shown in the previous section, these expressions do not depend on the 
position inside the cavity where the cavity fields are defined.
We finally deduce the mean radiation pressures on mirrors M1 and M2
\begin{eqnarray}
\left\langle P_{1}\right\rangle _{\mathrm{vac}}&=&
- \left\langle P_{2}\right\rangle _{\mathrm{vac}} \nonumber \\
&=& \sum_{m}\ \hbar \omega _{m} \cos ^{2}\theta _{m} \left( 1-g_{m}\right) 
\end{eqnarray}
At this point, it is worth emphasizing that we have assumed equilibrium at zero temperature 
for the whole system~: not only the input fields but also any fluctuations associated with 
loss mechanisms inside the mirrors correspond to zero-point fluctuations, whatever their 
microscopic origin may be. Otherwise, the expression of the force discussed in the following
would be affected.

The pressures have opposite values on the two mirrors M1 and M2.
This entails that the global force exerted by vacuum upon the cavity
vanishes, in consistency with the translational invariance of vacuum. In the
following, we denote $F$ the Casimir force calculated for M1 when considering
the limit of a large area $A\gg L^{2}$ 
\begin{equation}
F=A\left\langle P_{1}\right\rangle _{\mathrm{vac}}=A\sum_{m}\ \hbar \omega
_{m}\ \cos ^{2}\theta _{m}\ \left( 1-g_{m}\right)
\end{equation}
The sign conventions used here are such that the positive value obtained below for $F$ 
corresponds to an attraction of the two mirrors to each other.

\subsection{The force as an integral over real frequencies}

We now perform a change of variable to rewrite the summation symbol as specified in
(\ref{mphi}) 
\begin{eqnarray}
F&=&A\left\langle P_{1}\right\rangle _{\mathrm{vac}} \nonumber \\
&=&A\sum_{p}\int \frac{\mathrm{d}^{2} \mathbf{k}}{4\pi ^{2}} 
\int \frac{\mathrm{d}\omega }{2\pi } \hbar k_{z}\
\left( 1-g_{\mathbf{k}}^{p}\left[ \omega \right] \right)
\label{CasimirAiry}
\end{eqnarray}
We will now specify the domain of integration for $\omega $.

Up to now, we have discussed the scattering for ordinary waves which freely propagate 
in vacuum and correspond to frequencies $\omega$ larger than the bound 
$c\left| \mathbf{k}\right|$ fixed by the norm of the transverse wavevector. 
But we must also take into account the contribution of evanescent waves which 
correspond to frequencies $\omega$ smaller than $c\left| \mathbf{k}\right|$. 
These waves are fed by the additional fluctuations coming from the noise lines into 
the dielectric medium and propagating with an incidence angle larger than the limit angle. 
They are thus transformed at the interface into evanescent waves decreasing exponentially 
when the distance from the interface increases. 
As is well known \cite{BWevanescent}, the properties of these evanescent waves are 
conveniently described through an analytical continuation of those of ordinary waves.
This analytical continuation can only be dealt with in terms of functions having
a well defined analyticity behaviour. This is not the case for the Airy function 
$g_{\mathbf{k}}^{p}\left[ \omega \right]$ but we know that this function 
is the sum (\ref{gfunction}) of parts having well defined 
analyticity properties 
\begin{eqnarray}
g_{\mathbf{k}}^{p}\left[ \omega \right] &=&1+f_{\mathbf{k}}^{p}\left[ \omega \right]
+f_{\mathbf{k}}^{p}\left[ \omega \right] ^{\ast } 
=\frac{1-\left| \rho _{\mathbf{k}}^{p}\left[ \omega \right] \right| ^{2}}
{\left| 1-\rho _{\mathbf{k}}^{p}\left[\omega \right] \right| ^{2}}  \nonumber \\ 
f_{\mathbf{k}}^{p}\left[ \omega \right] &=&\frac{\rho _{\mathbf{k}}^{p}\left[ \omega \right] }
{1-\rho _{\mathbf{k}}^{p}\left[ \omega \right] } \nonumber \\
\rho _{\mathbf{k}}^{p}\left[ \omega \right] &=&r_{\mathbf{k},1}^{p}\left[ \omega \right] 
r_{\mathbf{k},2}^{p}\left[ \omega \right] e^{-2\kappa _{0}L}
\label{loopfunctions}
\end{eqnarray}
$\rho _{\mathbf{k}}^{p}\left[ \omega \right]$ is the `open loop function' corresponding
to one round trip of the field inside the cavity and defined as the product of the 
reflection amplitudes $r_{\mathbf{k},1}^{p}\left[\omega \right] $ and 
$r_{\mathbf{k},2}^{p}\left[ \omega \right] $ of the two mirrors and of the propagation 
phaseshift $e^{-2\kappa _{0}L}$; it is an analytical function in the physical domain 
of complex frequencies $\Re\xi > 0$ with the branch of the square root chosen so that 
$\Re\kappa >0$. Since the transverse wavevector is spectator 
throughout the whole scattering process, analyticity is defined with $\mathbf{k}$ fixed.

Then, $f_{\mathbf{k}}^{p}\left[ \omega \right] $ is the `closed loop function' built up 
on the open loop function $\rho _{\mathbf{k}}^{p}\left[ \omega \right]$. It is also an
analytical function, thanks to analyticity of the open loop and to a stability 
property which has a natural interpretation~: the system formed by the Fabry-Perot cavity 
and the vacuum fluctuations is stable because neither the mirrors nor the vacuum would have 
the ability to sustain an oscillation. In some cases, the stability can be derived from
a more stringent passivity property \cite{Lambrecht97} which may essentially be written
$\left| \rho _{\mathbf{k}}^{p}\left[ \omega \right] \right| <1$. 
However, the passivity property is sometimes too stringent to be obeyed by real mirrors
(see more detailed discussions in appendix \ref{AppEvan}). 
In any case, the stability property, {\it i.e.} the absence of self sustained 
oscillations, is sufficient for the present derivation of the Casimir force.

We are now able to give more precise specifications of the domain of integration in
(\ref{CasimirAiry}). Using the decomposition (\ref{loopfunctions}), we write the 
contribution of ordinary waves to this integral as the sum of two conjugated expressions
\begin{eqnarray}
F_{\mathrm{ord}}&=&\mathcal{F}_{\mathrm{ord}}+\mathcal{F}_{\mathrm{ord}}^{\ast } \nonumber \\
\mathcal{F}_{\mathrm{ord}} &=&-A\sum_{p}\int \frac{\mathrm{d}^{2}\mathbf{k}}{4\pi^{2}}
\int_{c\left| \mathbf{k}\right| }^{\infty }\frac{\mathrm{d}\omega }{2\pi }
\hbar k_{z}\ f_{\mathbf{k}}^{p}\left[ \omega \right] 
\end{eqnarray}
The integral $\mathcal{F}_{\mathrm{ord}}$ is built on the retarded function 
$f_{\mathbf{k}}^{p}\left[ \omega \right] $ which may be extended through an 
analytical continuation from the sector of ordinary waves to that of evanescent waves. 
The contribution of evanescent waves to the force is thus obtained as 
\begin{eqnarray}
F_{\mathrm{eva}}&=&\mathcal{F}_{\mathrm{eva}}+\mathcal{F}_{\mathrm{eva}}^{\ast } \nonumber \\
\mathcal{F}_{\mathrm{eva}}&=&-A\sum_{p}\int \frac{\mathrm{d}^{2}\mathbf{k}}{4\pi^{2}}
\int_{0}^{c\left| \mathbf{k}\right| }\frac{\mathrm{d}\omega }{2\pi } 
\hbar k_{z}\ f_{\mathbf{k}}^{p}\left[ \omega \right] 
\end{eqnarray}
The final expression of the Casimir force is the sum of the contributions of ordinary and 
evanescent waves that is also the integral over the whole axis of real frequencies 
\begin{eqnarray}
F&=&F_{\mathrm{ord}}+F_{\mathrm{eva}}=\mathcal{F}+\mathcal{F}^{\ast } \nonumber \\ 
\mathcal{F}&=&-A\sum_{p}\int \frac{\mathrm{d}^{2}\mathbf{k}}{4\pi ^{2}}
\int_{0}^{\infty }\frac{\mathrm{d}\omega }{2\pi } 
\hbar i\kappa_0 \ f_{\mathbf{k}}^{p}\left[ \omega \right] 
\label{ForceReal}
\end{eqnarray}
As far as ordinary waves are concerned, this corresponds to the intuitive picture 
where the Casimir force results from the radiation pressure of vacuum fluctuations 
filtered by the cavity \cite{Jaekel91}. The contribution of evanescent waves is but 
the extension of the domain of integration to the whole real axis with the cavity 
response function $f_{\mathbf{k}}^{p}\left[ \omega \right]$ extended through an 
analytical continuation. 
In the evanescent sector, the cavity function $f_{\mathbf{k}}^{p}\left[ \omega \right]$ 
is written in terms of reflection amplitudes calculated for evanescent waves and 
exponential factors corresponding to evanescent propagation through the cavity. 
This means that it describes the `frustration' of total reflection on one mirror due to 
the presence of the other. This explains why the radiation pressure of evanescent waves 
is not identical on the two sides of a given mirror and, therefore, how evanescent waves 
have a non null contribution to the Casimir force.

\subsection{The force as an integral over imaginary frequencies}

Using the Cauchy theorem, we now rewrite the Casimir force (\ref{ForceReal}) as an
integral over the axis of imaginary frequencies. 

\vspace*{4mm}\begin{figure}[tbh]
\centerline{\psfig{figure=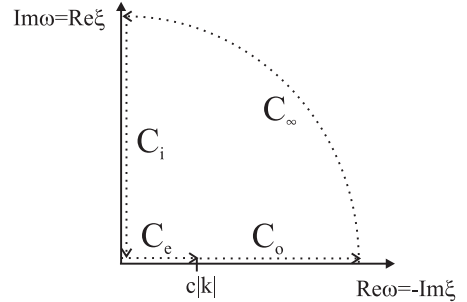,height=4cm}}
\caption{Contour representing the frequencies of interest for the evaluation 
of the Casimir force~: C$_\mathrm{o}$ and C$_\mathrm{e}$ correspond to the real
frequencies associated with ordinary and evanescent waves; C$_\mathrm{i}$ correspond
to the imaginary frequencies and C$_{\infty}$ to a quarter circle with a radius
allowed to go to infinity.}
\label{FigContour}
\end{figure}\vspace*{4mm}

Since $\kappa_0 f_{\mathbf{k}}^{p}\left[ i\xi \right] $ is analytical in the domain 
$\Re\xi >0$, its integral over a closed contour lying in this domain has to vanish. 
We choose the contour drawn on Figure \ref{FigContour} which consists of the positive 
part of the real axis including ordinary (C$_\mathrm{o}$) and evanescent 
(C$_\mathrm{e}$) waves, a quarter of circle C$_{\infty}$ with a very large radius and, 
finally, the imaginary axis C$_\mathrm{i}$ run from infinity to zero. 
Now the function $\kappa_0 f_{\mathbf{k}}^{p}\left[ i\xi \right] $ goes to zero for 
large values of the frequency, as a consequence of transparency at high frequency, 
a property certainly valid for any realistic model of optical mirror. 
Thanks to this property, the contribution to the integral of C$_{\infty}$ vanishes. 
We then deduce that the integrals over the real axis $\left[0,+\infty \right[$ 
and over the imaginary axis $\left[ 0,+i\infty \right[ $ are equal. 

We thus get a new expression of the force $F$ as an integral over imaginary 
frequencies $\omega$, that is also as an integral over real values of $\xi$, 
\begin{eqnarray}
F&=& \mathcal{F} +\mathcal{F}^{\ast } = 2\mathcal{F} \nonumber \\ 
\mathcal{F}&=&A\sum_{p}\int \frac{\mathrm{d}^{2}\mathbf{k}}{4\pi ^{2}}
\int_{0}^{\infty }\frac{\mathrm{d}\xi }{2\pi } 
\hbar\kappa_0 \ f_{\mathbf{k}}^{p}\left[ i\xi \right] \nonumber \\
&&\kappa _{0}=\sqrt{\mathbf{k}^{2}+ \frac{\xi ^{2}}{c^2}}  
\label{Force}
\end{eqnarray}
We have used the fact that $\mathcal{F}$ is real, so that $\mathcal{F}^{\ast }$ is 
simply equal to $\mathcal{F}$. This property is less obvious, but also true, with 
$\mathcal{F}$ written as an integral over real frequencies. 
We wish to emphasize more generally that expression (\ref{Force}) is 
mathematically equivalent to (\ref{ForceReal}). 
The former expression is closer to the physical intuition whereas 
the latter is better adapted to explicit computations of the force.

Expression (\ref{Force}) gives the Casimir force between real mirrors described by 
arbitrary frequency dependent reflection amplitudes. It is a regular integral as 
soon as these amplitudes obey the physical assumptions used in the derivation~: 
causality, unitarity and high frequency transparency for each mirror, 
stability of the system formed by the two mirrors and the scattered vacuum fields. 
The demonstration holds for dissipative mirrors and not only for lossless ones.  

The limit of perfect mirrors is obtained in expression (\ref{Force}) by letting 
the reflection amplitudes go to unity, which leads to the Casimir formula 
(\ref{eqCasimir}). 
This can be considered as an alternative demonstration of the Casimir formula 
without any reference to a renormalization or regularization technique.
Basically, the properties of real mirrors, in particular their high frequency 
transparency, are sufficient to provide a regular expression
of the force, as it was guessed a long time ago by Casimir \cite{Casimir48}.

As a simple model of the mirrors used in the experiments, let us consider a 
metallic slab with a large width, that is a width $\ell $ larger than a few plasma 
wavelengthes. 
We use expression (\ref{Force}) of the force written as an integral over imaginary
values of the frequencies ($\omega=i\xi$, $\xi$ real). Hence, the phase factor  
corresponding to one round trip inside the slab is a decreasing exponential
with a real exponent $e^{-2\kappa_1\ell}$. 
For the plasma model (\ref{PlasmaModel}), $\kappa_1$ is given by 
$\sqrt{ \frac{\xi^2}{c^2}+\frac{\omega_\mathrm{P}^2}{c^2}+\mathbf{k}^2}$
and it is larger than $\frac{2\pi}{\lambda_\mathrm{P}}$ for all values
of $\xi$ and $\mathbf{k}$.
When relaxation is taken into account, this is still the case
except in a very narrow domain with values of $\xi$ and $\mathbf{k}$
both close to zero. This domain has a negligible contribution to the
integral (\ref{Force}) and it follows that the reflection amplitude of the slab may 
be replaced by the limiting expression obtained for the bulk. One thus recovers the 
Lifshitz expression for the Casimir force \cite{Lifshitz56} which is widely used for
comparing experimental results with theoretical expectations \cite{Bordag01}.

\subsection{Constraints on the force}

We now deduce general constraints which invalidate proposals 
made for tayloring the Casimir force at will by using specially designed mirrors 
\cite{Iacopini93,Ford93}. This generalizes to 3D space the results obtained for 
1D space in \cite{Lambrecht97} to which the reader is referred for further discussions.

Expression (\ref{Force}) is an integral over the axis of imaginary frequencies 
essentially determined by the reflection amplitudes 
$r_1\left[ i\xi \right]$ and $r_2\left[ i\xi \right]$ for $\xi$ real. 
These amplitudes always have a modulus smaller than unity, for arbitrary dielectric 
multilayers (see appendix \ref{AppPassiv}).
They are negative for arbitrary dielectric slabs (see appendix \ref{AppSlab}) and 
we deduce from the composition law (\ref{compoSlab}) that this is still the case  
for arbitrary dielectric multilayers. 
It follows that the product of the reflection amplitudes of the two mirrors 
is always positive with a modulus smaller than unity 
\begin{equation}
0<r_{1}\left[ i\xi \right] r_{2}\left[ i\xi \right] <1
\end{equation}

From this, we deduce first that the Casimir force has an absolute value
smaller than the value (\ref{eqCasimir}) reached for perfect mirrors and 
that it remains attractive 
\begin{equation}
0\leq F\leq F_{\mathrm{Cas}}
\end{equation}
We also derive that the Casimir force decreases as a function of the length 
\begin{equation}
\frac{\mathrm{d}F}{\mathrm{d}L}\leq 0
\end{equation}
This means that the properties obeyed by real mirrors strongly constrain 
the possible variation of the Casimir force, contrarily to what
might have been expected at first sight \cite{Iacopini93,Ford93}.  

Note that we have considered mirrors used in the experiments which
have electric permittivity but no magnetic permeability. 
Different results would be obtained with magnetic mirrors, precisely with 
one of the mirrors dominated by electric response and the other one by 
magnetic response. The product of the two reflection amplitudes would 
indeed be negative in this case and the Casimir force repulsive
\cite{Boyer74,Kupiszewska93,Alves00,Kenneth02}. 

\section{Conclusion}

We have presented a derivation of the Casimir force between lossy mirrors 
characterized by arbitrary frequency dependent reflection amplitudes,
in the Casimir geometry where the cavity is made with two parallel plane mirrors. 

We have shown how mirrors and cavities may be dealt with by using a quantum theory of 
optical networks. We have deduced the additional fluctuations accompanying dissipation 
from expressions of the optical theorem adapted to quantum network theory. 
The optical theorem is equivalent to the unitarity of the whole scattering process
which couples the modes of interest and the noise modes and it ensures that 
the quantum commutators of the output fields are the same as those of the input fields. 
The situation is different for the cavity fields which do not freely propagate. 
We have given a general proof of a theorem previously demonstrated in particular 
cases \cite{Jaekel91,Barnett98} which states that the modification of the commutators 
is determined by the usual Airy function, that is the spectral density associated
with the Fabry-Perot cavity. For arbitrary lossy mirrors, the spectral density 
is determined by the reflection amplitudes as they are seen by the intracavity fields.  
It determines the radiation pressure exerted by vacuum fluctuations 
upon the mirrors with repulsive and attractive contributions associated respectively 
with resonant or antiresonant frequencies. The Casimir force is then obtained as an 
integral over the whole axis of real frequencies, including the contribution of
evanescent waves besides that of ordinary waves. It is equivalently expressed as an 
integral over imaginary frequencies. The derivation only uses a few general assumptions 
certainly valid for real optical mirrors, namely causality, unitarity, 
high-frequency transparency for each mirror and stability of the compound cavity-vacuum 
system. It leads to a finite result without any further reference to a regularization
technique \cite{Jaekel91}. 

The formula obtained in the present paper for the Casimir force was already known
\cite{Jaekel91} but its scope of validity is widened by the
present demonstration. It has been used to discuss the effect of imperfect
reflection for the metallic mirrors used in the experiments. Different descriptions
of the optical response of metals have been used, from the crude application 
of the plasma model (\ref{PlasmaModel}) to a more complete characterization of the 
dielectric constant derived from tabulated optical data and dispersion relations. 
This kind of calculations, discussed in great detail for the mirrors 
corresponding to the recent experiments (see for example \cite{Lambrecht00}),
has not been reproduced here.
Instead, we have presented general results valid for any real mirrors obeying 
the physical properties already evoked and shown that they strongly constrain 
the variation of the Casimir force.

In the present paper, we have restricted our attention on the limit of zero 
temperature although our work was partly motivated by a recent polemical
discussion of the effect of temperature on the Casimir force between real
mirrors \cite{Bostrom00,Svetovoy00,Bordag00,Lamoreaux01c,Sernelius01r,%
Sernelius01c,Bordag01r,Klimchitskaya01,Bezerra02,Lamoreaux02}. 
Since contradictory results may have raised doubts about the validity and
consistency of various derivations of the Casimir force, we have considered 
it was important to come back to the first principles in this derivation. 
This has been done in the present paper for the case of zero temperature. 
A follow-on publication will show how to include the effect of thermal fluctuations 
in the treatment in order to obtain an expression free from ambiguities for the
Casimir force between arbitrary lossy mirrors at non zero temperatures.

\appendix

\section{The dielectric slab}
\label{AppSlab}

In this appendix, we discuss in more detail the specific
case of the dielectric slab. We consider lossy as well as lossless slabs.

For a lossless dielectric medium, the permittivity $\varepsilon $ is real at real
frequencies. For ordinary waves, $\kappa _{0}$ and $\kappa _{1}$
are purely imaginary, so that the impedance ratios are real for both polarizations. 
Hence $\beta$ is real ($\beta=\beta_r$) and $\alpha$ purely imaginary 
($\alpha=i\alpha_i$) so that the scattering amplitudes (\ref{slab}) are read as 
\begin{eqnarray}
t &=&\frac{\sinh \beta _{r}}{\sinh \left( \beta _{r}+i\alpha _{i}\right) } \nonumber \\
&=&\frac{\sinh \beta _{r}}{\sinh \beta _{r}\cos \alpha _{i}+i\cosh \beta
_{r}\sin \alpha _{i}}  \nonumber \\
r &=&-\frac{\sinh \left( i\alpha _{i}\right) }{\sinh \left( \beta
_{r}+i\alpha _{i}\right) } \nonumber \\
&=&-\frac{i\sin \alpha _{i}}{\sinh \beta _{r}\cos
\alpha _{i}+i\cosh \beta _{r}\sin \alpha _{i}}
\end{eqnarray}
The sum of the squared amplitudes is unity 
$\left| t\right| ^{2}+\left| r\right| ^{2}=1 $ while the reflexion and transmission 
amplitudes are in quadrature to each other $tr^{\ast }+rt^{\ast }=0$, which means 
that $S$ is a unitary $2\times 2$ matrix, as it was expected for a lossless mirror. 
This implies that the reflection amplitude has a modulus smaller than unity 
$\left| r\right| <1$. This property also holds for lossy mirrors thanks to
positivity of dissipation (see appendix \ref{AppPassiv}).

Unitarity is defined without ambiguity only in the case of ordinary waves.
For a lossless slab and evanescent waves, $\kappa_0$ is real - it is just the 
inverse of the penetration length of evanescent wave in vacuum - whereas 
$\kappa _1$ remains purely imaginary. 
Hence $\beta$ as well as $\alpha$ are purely imaginary and it is no longer 
possible to obtain general bounds for the scattering amplitudes 
\begin{eqnarray}
t&=&\frac{\sinh \left( i\beta _{i}\right) }{\sinh \left( i\beta _{i}+i\alpha
_{i}\right) }=\frac{\sin \beta _{i}}{\sin \left( \beta _{i}+\alpha
_{i}\right) }\nonumber \\ 
r&=&-\frac{\sinh \left( i\alpha _{i}\right) }
{\sinh \left( i\beta _{i}+i\alpha _{i}\right) }=-\frac{\sin \alpha _{i}}
{\sin\left( \beta _{i}+\alpha _{i}\right) }
\end{eqnarray}
In particular, $\left| r\right|$ does not remain always smaller than 1
(see more explicit discussions in appendix \ref{AppEvan}
with different results for the TE and TM polarizations). 

For imaginary frequencies finally, $\beta $ and $\alpha $ are positive real 
numbers, for lossy as well as lossless slabs. 
In this case, general bounds are easily obtained for the amplitudes 
\begin{eqnarray}
&&0<t=\frac{\sinh \left( \beta _{r}\right) }{\sinh \left( \beta _{r}+\alpha
_{r}\right) }<1 \nonumber \\ 
&&0<-r=\frac{\sinh \left( \alpha _{r}\right) }
{\sinh \left( \beta _{r}+\alpha _{r}\right) }<1
\label{boundRTimaginary}
\end{eqnarray}
The fact that $r$ is negative with a modulus smaller than unity plays an important
role in the derivation of constraints on the Casimir force.

Interesting results are also obtained for the eigenvalues of the $S-$matrix, which 
have a simple form $s_\pm = r \pm t$ since the slab is symmetrical in the exchange of
its two ports. In the sector of ordinary waves, unitarity 
(\ref{unitarity}) has a simple form in terms of $s_\pm = r \pm t$ and of the similar
quantities $s_\pm^\prime = r^\prime \pm t^\prime $ defined on the noise matrix $S^\prime $ 
\begin{equation}
\left| s_\pm \right| ^{2}+\left| s_\pm ^\prime \right| ^{2}=1
\label{unitaritySSprime}
\end{equation} 
For the lossless slab, $s_\pm$ have a unit modulus and $s_\pm ^\prime$ vanish. For a lossy
slab, we have 
\begin{equation}
\left| s_\pm \right| ^{2} \leq 1
\label{boundS}
\end{equation} 
This can be considered as a consequence of (\ref{unitaritySSprime}) with 
$\left| s_\pm^\prime \right| ^{2} \geq 0$. Equivalently, it can be considered that
unitarity (\ref{unitaritySSprime}) fixes the modulus of $s_\pm ^\prime$ when the modulus
of $s_\pm$ is known. 

Condition (\ref{boundS}) will be found in appendix \ref{AppPassiv} to express
a passivity property for the slab, here for ordinary waves. This property still holds 
in the sector of imaginary frequencies, as a consequence of (\ref{boundRTimaginary}) 
and of the following inequalities obeyed for all positive real numbers $\alpha$ 
and $\beta$  
\begin{eqnarray}
\left| \sinh \beta \mp \sinh \alpha \right| \leq {\sinh \left( \alpha +\beta
\right) }  
\end{eqnarray}
Using the terms of appendix \ref{AppPassiv}, this means that the domain
of passivity always includes the sectors of ordinary waves and imaginary frequencies,
in the case of a dielectric slab. However, it does not necessarily include the sector
of evanescent waves (see appendix \ref{AppEvan}).

\section{The sector of evanescent waves}
\label{AppEvan}

Ordinary waves correspond to frequencies $\omega \geq c\left| \mathbf{k}\right|$ 
and real wavevectors $k_{z}$ whereas evanescent waves correspond to frequencies 
$\omega \leq c\left| \mathbf{k}\right|$ and imaginary values of $k_{z}$. 
Causal scattering amplitudes can be extended from ordinary to evanescent waves, 
by an analytical continuation through the physical domain of complex frequencies 
$\omega=i\xi$ with $\Re\xi > 0$ and $\Re\kappa >0$. 
The `energy conditions' which bear on quadratic forms are not necessarily 
preserved in this process. 

In order to illustrate the idea, let us consider the reflection amplitude (\ref{r01}) at 
the interface between vacuum ($\varepsilon _0=1$) and a lossless dielectric medium 
($\varepsilon _1$ real for $\omega$ real). In the sector of evanescent waves, $\kappa _1$ 
is imaginary and $\kappa _0$ real, so that $r$ and $\overline{r}$ are complex numbers 
with a unit modulus, that is also pure dephasings corresponding to the phenomenon
of total reflection. Meanwhile, the transmission amplitudes differ from zero, which describes
how evanescent waves in vacuum are fed by the fields coming from the dielectric medium 
with an incidence angle larger than the limit angle. In these conditions, it is clear that 
the condition $\left| r\right| ^{2}+\left| t\right| ^{2}\leq 1$ fails. 

For the TE polarization, it turns out that 
\begin{eqnarray}
\left| r^\mathrm{TE}\right| \leq 1
\label{rTEeva}
\end{eqnarray}
in the evanescent sector at the interface between vacuum and any dielectric medium. 
This property is always true in the sectors
of ordinary waves and imaginary frequencies for an arbitrary mirror (see the appendices
\ref{AppSlab} and \ref{AppPassiv}). Using high frequency transparency, it follows from the 
Phragm\'en-Lindel\"of theorem \cite{Phragmen} that inequality (\ref{rTEeva}) holds in the 
whole physical domain in the complex plane. This ensures that the closed loop 
function $f ^\mathrm{TE}$ is analytic and, in particular, has no pole in the domain 
$\Re\xi >0$. In other words, since the open loop gain is smaller than unity, 
the closed loop cannot reach the oscillation threshold, leading to the stability
property used in the derivation of the Casimir force. 

Although it seems quite natural, this argument is 
not valid in the general case. For metallic mirrors for example, the condition 
$\left| r \right| \leq 1$ is violated in the evanescent sector for TM modes. 
The reflection amplitude is even known to reach large resonant values at the 
plasmon resonances \cite{Barton79}. Of course, this does not prevent the stability 
property to be fulfilled~: the Fabry-Perot cavity is in this 
case a stable closed loop built on an open loop exceeding the unit modulus but with a 
phase such that the oscillation threshold is not reached. 

We stress again that the stability property is necessary
in the derivation of the Casimir force since it entails 
that the closed loop function is properly defined in the evanescent sector. 
When the more stringent property $\left| r \right| \leq 1$ is also obeyed, it follows
from expression (\ref{loopfunctions}) that the Airy function, which has been defined 
with the significance of a positive spectral density on ordinary waves, remains positive 
in the evanescent sector. When the property $\left| r \right| \leq 1$ fails, the 
Airy function can no longer be thought of as a spectral density in the whole physical
domain, but this does not invalidate the derivation of the Casimir force.

\section{The domain of passivity}
\label{AppPassiv}

In this appendix, we discuss the related but not identical properties corresponding to 
positivity of dissipation and passivity. 

We consider an arbitrary mirror, that is a reciprocal network connecting two vacuum ports. 
For ordinary waves, we define the power dissipated by the mirror 
\begin{eqnarray}
\pi &=&\left( e_{\mathrm{L}}^{\mathrm{in}} {}^\dagger e_{\mathrm{L}}^{\mathrm{in}}
-e_{\mathrm{L}}^{\mathrm{out}} {}^\dagger e_{\mathrm{L}}^{\mathrm{out}} \right) 
+\left( e_{\mathrm{R}}^{\mathrm{in}} {}^\dagger e_{\mathrm{R}}^{\mathrm{in}}
-e_{\mathrm{R}}^{\mathrm{out}} {}^\dagger e_{\mathrm{R}}^{\mathrm{out}}\right)  
\nonumber \\
&=&\left\langle e^{\mathrm{in}}\right| \left| e^{\mathrm{in}}\right\rangle
-\left\langle e^{\mathrm{out}}\right| \left| e^{\mathrm{out}}\right\rangle
\end{eqnarray}
where we have introduced row vectors conjugated to the column vectors 
\begin{equation}
\left\langle e^{\mathrm{out}}\right| =\left| e^{\mathrm{out}}\right\rangle
^\dagger \qquad \qquad \left\langle e^{\mathrm{in}}\right| =\left| 
e^{\mathrm{in}}\right\rangle ^\dagger 
\end{equation}
This power is positive as a consequence of unitarity
\begin{eqnarray}
\pi &=& \left\langle e^{\mathrm{in}}\right| I-S^\dagger S\left| 
e^{\mathrm{in}}\right\rangle \nonumber \\
&=& \left\langle e^{\mathrm{in}}\right| 
S^{\prime \ \dagger}S^\prime \left| e^{\mathrm{in}}\right\rangle \geq 0
\end{eqnarray}
which corresponds to the positivity of the matrix $I- S^\dagger S$ 
\begin{equation}
\forall \left| e\right\rangle \qquad \qquad \left\langle e\right|
I-S^\dagger S\left| e\right\rangle \geq 0
\label{defPassiv}
\end{equation}
where $\left| e\right\rangle $ represents arbitrary input fields.
Positivity can also be expressed in terms of the eigenvalues $\ell $ of
$S^\dagger S$ 
\begin{equation}
\det \left( S^\dagger S-\ell I\right) =0\qquad \qquad \ell \geq 0
\end{equation}
These eigenvalues are always real and positivity of dissipation is equivalent 
to the fact that they are smaller than unity 
\begin{equation}
\ell \leq 1
\label{def2Passiv}
\end{equation}

Passivity is a property directly related to positivity of dissipation but 
defined more generally for complex frequencies in the physical domain. 
In order to discuss it,
we extend the matrix $S$ from the sector of ordinary waves through the analytical
continuation already discussed. We extend $S^\dagger $ similarly,
with the complex conjugation cautiously defined since it involves complex frequencies~:
conjugation corresponds to $\xi \rightarrow \xi ^\ast$ and $\kappa \rightarrow \kappa ^\ast$ 
and it preserves the physical domain $\Re\xi >0\ ,\ \Re\kappa >0$; the derivations performed 
for an amplitude in the domain $\Re\xi >0\ ,\ \Im\xi <0$ are thus translated to similar 
derivations for the conjugated amplitude in the quarter plane $\Re\xi >0\ ,\ \Im\xi >0$.

Then, the domain of passivity of $S$ is defined by the domain of $\xi$ for which 
$I-S^\dagger S$ is a positive matrix (eq.\ref{defPassiv}) that is also 
for which the eigenvalues $\ell$ of $S^\dagger S$ are smaller than unity 
(eq.\ref{def2Passiv}).
An important feature of this property is that it is stable under composition~: 
when two networks A and B are piled up as in Figure \ref{FigComposition}, the 
quadratic forms appearing in (\ref{defPassiv}) simply add up so that passivity of the 
network AB follows from passivity of the two networks A and B. This is a special case 
of a general theorem \cite{Meixner} 
which states that networks built up with passive elements are passive. 

Passivity means that the eigenvalues $1-\ell$ of the matrix $I-S^\dagger S$ are both 
positive, which is equivalent to the following inequalities 
\begin{equation}
\mathrm{Tr}\left( I-S^\dagger S\right) \geq 0 \qquad 
\det \left(I-S^\dagger S\right) \geq 0
\end{equation}
It may be written in terms of the scattering amplitudes 
\begin{eqnarray}
&&\left| r\right| ^{2}+\left| \overline{r}\right| ^{2}+2\left| t\right|
^{2}\leq 2\nonumber \\ &&\left| r\overline{r}-t^{2}\right| ^{2}\geq \left|
r\right| ^{2}+\left| \overline{r}\right| ^{2}+2\left| t\right| ^{2}-1
\end{eqnarray}
Passivity implies that the scattering amplitudes have a modulus
smaller than unity 
\begin{equation}
\left| r\right| \leq 1 \qquad \left| \overline{r}\right| \leq 1 \qquad
\left| t\right| \leq 1
\end{equation}
Conversely, the latter conditions are necessary but not sufficient 
for passivity.

For a mirror symmetrical in the exchange of its two ports, a slab for example, the 
passivity conditions take the simple form $\left| r \pm t \right| ^{2} \leq 1$.
The results of appendix \ref{AppSlab} thus entail that the domain of passivity 
always includes the sectors of ordinary waves and imaginary frequencies, 
for arbitrary slabs (eq.\ref{boundS}). 
Using the stability of passivity under composition, we deduce that this is also the 
case for arbitrary multilayers. It follows that the reflection amplitudes always 
have a modulus smaller than unity for imaginary frequencies ($\xi$ real)
\begin{equation}
\left| r\left[ i\xi \right] \right| \leq 1 
\label{passivImaginary}
\end{equation}

\medskip

\begin{acknowledgments}
Thanks are due to Gabriel Barton and Marc-Thierry Jaekel for their helpful comments.
\end{acknowledgments}

\end{document}